\documentclass[trackchanges,twocolumn,resetfootnote]{aastex701}
\usepackage{CJKutf8} 
\usepackage{amsmath}
\usepackage{comment}
\usepackage{physics}
\usepackage{textcomp}
\usepackage{multirow}
\usepackage{soul}
\usepackage{threeparttable}

\newcommand\kms{km$\,$s$^{-1}$}
\newcommand\Msol{M$_{\odot}$}

\newcommand{\hi}{H\,{\sc i}}
\newcommand{\re}{$r_{\rm e}$}
\newcommand{\ngc}{$N_{\rm GC}$ }

\shortauthors{Khim et al.}
\shorttitle{Clusters in SMUDGes}
\received{August 24, 2026}
\submitjournal{ApJ}

\begin{document}

\title{SMUDGes Below the Surface: Clusters Confirm the Diversity of Low-Surface-Brightness Galaxies}

\author[orcid=0000-0002-7013-4392]{Donghyeon J. Khim}
\affiliation{Steward Observatory and Department of Astronomy, University of Arizona, 933 N. Cherry Avenue, Tucson, AZ 85721, USA} 
\email[show]{galaxydiver@arizona.edu}

\author[orcid=0000-0002-5177-727X]{Dennis Zaritsky}
\affiliation{Steward Observatory and Department of Astronomy, University of Arizona, 933 N. Cherry Avenue, Tucson, AZ 85721, USA} 
\email{dfz@arizona.edu}

\author[orcid=0000-0001-8568-8729]{Loraine Sandoval Ascencio} 
\affiliation{Department of Physics \& Astronomy, University of California, Irvine, 4129 Reines Hall, Irvine, CA 92697, USA} 
\email{lorainas@uci.edu}

\author[orcid=0000-0003-1371-6019]{M. C. Cooper}
\affiliation{Department of Physics \& Astronomy, University of California, Irvine, 4129 Reines Hall, Irvine, CA 92697, USA} 
\email{cooper@uci.edu}

\author[orcid=0000-0001-7618-8212]{Richard Donnerstein} 
\affiliation{Steward Observatory and Department of Astronomy, University of Arizona, 933 N. Cherry Avenue, Tucson, AZ 85721, USA}
\email{rdonnerst@gmail.com}

\thanks{Corresponding author: \href{mailto:galaxydiver@arizoan.edu}{galaxydiver@arizona.edu}}

\begin{abstract}

We use spatially resolved Keck Cosmic Web Imager spectroscopy to study high-surface-brightness stellar components and internal kinematics in 44 low-surface-brightness galaxies from the SMUDGes survey, including 30 ultradiffuse galaxies, spanning the red sequence, green valley, and blue cloud regions of the galaxy color-magnitude diagram. Although these galaxies appear largely smooth in survey imaging, we detect at least one cluster candidate in every galaxy. We identify clusters with nebular emission lines in 24 galaxies. They are common in blue-cloud galaxies but rare in red sequence galaxies. We identify three nuclear star cluster candidates, one of which shows nebular emission lines. Among clusters without nebular emission lines, bluer galaxies tend to have brighter cluster luminosity functions, suggesting the presence of young or intermediate-age clusters in addition to old globular clusters. This age diversity may complicate the use of photometrically selected globular cluster populations as tracers of total halo mass. We measure rotation velocities for 17 galaxies using the clusters with nebular emission lines and for 12 using diffuse stellar light. Within 0.5 effective radii, the inferred dynamical mass generally exceeds the stellar mass, although the inferred dark matter contribution varies by an order of magnitude. The variety in properties for this sample of ultradiffuse galaxies and low-surface-brightness dwarfs suggests that no single formation mechanism explains the population of low-surface-brightness galaxies.

\end{abstract}




\section{Introduction} 
\label{sec:intro}

Ultradiffuse galaxies (UDGs) have stellar masses comparable to those of dwarf galaxies but unusually extended stellar distributions. They are defined by effective radii $> 1.5$ kpc with central $g$-band surface brightness $ \ge 24$ mag arcsec$^{-2}$ \citep{2015vanDokkum}. 
Their large physical sizes and low stellar densities have motivated a broad range of proposed formation pathways, including feedback-driven expansion, high-angular-momentum halos, tidal processing, mergers, and environmental transformation \citep[e.g.,][]{Yozin_2015,Amorisco2016,DiCintio2017,Chan2018,Carleton2019,Jiang2019,Sales_2020,Wright2021,Benavides_2023}. 
Because size and surface brightness alone define UDGs, the category may include galaxies with distinct properties and evolutionary histories.
Identifying the full range of the variety among these enigmatic galaxies may help distinguish between different evolutionary pathways.

Observations have consistently hinted at significant variation in the UDG population. UDGs span broad ranges in color, gas content, stellar population, and environment \citep[e.g.,][]{Buzzo2025}. The population includes red-quenched galaxies and blue-star-forming galaxies. Redder UDGs are more prevalent in dense environments, whereas bluer and actively star-forming UDGs occur more frequently in the field \citep[e.g.,][]{2015vanDokkum,koda,prole+18, greco, leisman,kadowaki21}. Stellar population studies likewise indicate that UDGs can have substantially different star formation and assembly histories \citep[e.g.,][]{Ferre2018,Ferre2023,forbes24,loraine,SMUDGes9}.
This diversity is consistent with the possibility that multiple formation pathways contribute to the UDG population.

Compact stellar components provide another way to investigate this diversity. UDGs host globular cluster (GC) systems ranging from sparse to unusually rich \citep[e.g.,][]{beasley2016,Amorisco2018,Lim2018,Toloba_2018,marleau2024}, and some host nuclear star clusters \citep[NSCs,][]{Lim2018,lim,lambert,khim+24_nsc}. Field UDGs can host younger clusters or knots with nebular emission that trace localized recent star formation \citep[e.g.,][]{Trujillo2017,Khim+25}. Because these components have higher surface brightness than the diffuse stellar body, their luminosities, spatial distributions, and spectra can often be measured even when the underlying galaxy remains difficult to characterize in detail.

These components trace different stages of galaxy evolution. Old GCs preserve information about early star formation and halo assembly, whereas young clusters and emission-line regions trace ongoing or recent star formation. NSCs may grow through cluster migration \citep[e.g.,][]{tremaine1975, lotz, CDMB2009}, in-situ star formation \citep[e.g.,][]{Bailey1980, mihos94, bekki01, Seth2006}, or a combination of both. Differences in cluster abundance, luminosities, spatial distributions, and kinematics can therefore reveal whether similarly diffuse galaxies experienced similar or distinct evolutionary histories. In particular, unusually luminous GC-like objects may indicate younger or intermediate-age populations rather than conventional old GCs, linking the present-day cluster population to the recent evolution of the host galaxy.

Compact stellar components can also help us identify the reason for the large apparent range of dark matter properties. 
In some UDGs, kinematics indicates substantial dark matter contributions within one effective radius  \citep[e.g.,][]{vdk19,Gannon2020,forbes+21}. 
In other UDGs, dynamical masses are comparable to the enclosed stellar masses, suggesting that these are dark-matter-deficient galaxies \citep{vdk18_Gc_in_DM_free,vdk19_dmfree,Danieli2019,Pina2019,Pina+2020}. 
Compact stellar populations, with their higher surface brightness, provide opportunities for spectroscopic measurements for a larger sample of galaxies.

Field UDGs remain relatively understudied, in part because their low surface brightness makes spectroscopic confirmation and distance measurements challenging, while their sparse distribution across the sky makes systematic follow-up less efficient than for cluster environments. Many retain gas and ongoing or recent star formation, and some host localized high-surface-brightness regions with strong nebular emission \citep[e.g.,][]{leisman,karunakaran}. 
Deep integral-field spectroscopy provides the combination of sensitivity, spatial resolution, and spectral information needed to characterize UDGs within a single data set. Nebular emission lines identify localized recent star formation and provide precise velocities for emission-line clusters, while stellar absorption features trace the diffuse stellar body. The same observations can identify GC candidates, confirm cluster membership when sufficient spectral signal is available, and constrain the internal dynamical masses.  Such observations can simultaneously quantify the variations in the populations of compact sources in these galaxies and also provide new kinematic measurements to explore the distribution of mass.

Our previous study of the ``Disco Ball'', SMDG0038365-064207, demonstrated the value of combining these diagnostics \citep{Khim+25}. This galaxy lies in the green valley of the color--magnitude diagram and hosts an NSC, numerous clusters, and several clusters with narrow nebular emission lines. We also detected measurable rotation in both the diffuse stellar absorption and the clusters with emission lines. Its clusters appeared brighter than typical old GCs, suggesting that some of the clusters may have intermediate ages. The Disco Ball therefore linked cluster evolution, recent star formation, internal kinematics, and halo mass inference within a single system.

In this work, we extend the analysis to the full sample of 44 low-surface-brightness (LSB) galaxies observed with the Keck Cosmic Web Imager as part of the SMUDGes spectroscopic follow-up program. The sample spans the red sequence, green valley, and blue cloud. Following our analysis of the Disco Ball, we identify high-surface-brightness stellar components throughout the sample and examine how their populations vary with host galaxy color. To investigate internal kinematics, we measure velocity gradients using clusters with nebular emission lines and stellar absorption features in the galaxy light. We also estimate dynamical masses and infer halo masses from the GC populations to constrain the galaxies' dark matter content. Through these measurements, we ask whether galaxies that appear similarly diffuse in survey imaging share comparable internal structures and evolutionary states, and whether other field LSB galaxies host the same combination of recent cluster formation, unusually luminous non-emission-line clusters, and ordered internal rotation found in the Disco Ball.

The paper is organized as follows. In \S\ref{sec:method}, we describe the acquisition and reduction of KCWI data, including cluster finding and spectral analysis. In \S\ref{sec:results}, we present the cluster population and kinematics. In \S\ref{sec:discussion}, we will discuss the implications of our measurement. We summarize our main findings in \S\ref{sec:summary}. We assume the WMAP9 cosmological parameters throughout the analysis \citep{wmap9}, but variations within the currently allowed range of cosmological parameters do not materially affect our results. We report all magnitudes in the AB system \citep{oke1,oke2}.

\section{Methodology}
\label{sec:method}

\subsection{Data Acquisition and Reduction}
\label{sec:reduction}

The galaxies studied here were originally identified as UDG candidates in the SMUDGes survey \citep{smudges, smudges5}, which cataloged thousands of LSB galaxies and enabled statistical studies of the UDG population across all environments. As part of a spectroscopic follow-up program, \cite{loraine} observed 44 SMUDGes galaxies located outside of galaxy clusters, with the goal of investigating their recent star formation activity, using the Keck Cosmic Web Imager (KCWI) on Keck II \citep{kcwi_martin, kcwi}. From this sample, we previously focused on a single object, the ``Disco Ball'', a particularly noteworthy system because of its relatively red integrated color but the clear presence of numerous apparent clusters, including star-forming regions, in the KCWI data \citep{Khim+25}. In the present work, we extend the analysis of clumps/clusters to the full set of KCWI-SMUDGes galaxies.

We observed the targets between fall 2021 and fall 2022 with the KCWI medium image slicer in combination with the BL grating. We centered the instrumental setup at 4500 \AA\, which provides spectral coverage from 3500 to 5500 \AA\ and a field of view of 16 $\times$ 20$^{\prime\prime}$. 
The total integration time per target ranges from 1200 to 4800 seconds, divided into 2 to 8 individual exposures. To enable accurate sky subtraction, we obtained at least one sky exposure for non-local sky subtraction. We offset these sky exposures by approximately 60$^{\prime\prime}$ from each target, with exposure times typically matched to that of an individual science frame. In addition, we collected standard calibration frames, including bias, arc, and flat-field exposures. Across our sample, the KCWI field of view covers out to radii of $0.29$--$2.02$ effective radii (\re), with a median coverage out to $0.93$ \re. 

We reduced the data with the KCWI Data Reduction Pipeline\footnote{\url{https://kcwi-drp.readthedocs.io/en/latest/}}. \cite{loraine} provide a detailed description of the observing strategy and reduction procedure. For most targets, we performed non-local sky subtraction using the separate offset sky frames. For 11 targets, however, local sky subtraction produced higher signal-to-noise spectra, so we adopted those reductions.

As discussed in our ``Disco Ball'' paper, some pixels remain affected by cosmic rays even after applying the standard reduction procedures, including cosmic-ray rejection. To further mitigate these residual artifacts, we apply an additional filtering procedure based on statistical outlier rejection across exposures. Specifically, in a given exposure image, we identify outlier pixel values that both lie outside the 1st--99th percentile range and deviate by more than 3$\sigma$ from the median. We then compare these candidate pixels with the corresponding pixels in the other exposures and reject values that are inconsistent with the ensemble distribution, defined here as those falling outside the 5th--95th percentile range.
The rejected pixels occupy localized regions in both the spatial and spectral dimensions, as expected for residual cosmic-ray contamination.

\subsection{The Point and Spectral Line Spread Functions}
\label{sec:quality}

To ensure consistency with our previous analysis, we adopt the same characterization of the point spread function (PSF) and instrumental line broadening (ILB) derived from the Disco Ball dataset. Briefly, we treat the nuclear star cluster (NSC) in the Disco Ball as an unresolved source, given its expected physical size ( $<$ 40 pc, corresponding to $<$ 0.18$^{\prime\prime}$; \citealt{Georgiev2016, neumayer}). We therefore use the NSC as an empirical probe of the PSF, as it is the brightest point source in that field. A two-dimensional Gaussian profile is fitted independently along the RA (Right Ascension) and Dec (Declination) directions to account for the anisotropic plate scale of KCWI. From these fits, we measure full-width half maximum (FWHM) values of 1.23 and 2.87 spaxels in the RA and Dec directions, respectively, corresponding to an angular resolution of 0.84 $\pm$ 0.02$^{\prime\prime}$. This result is consistent with the expected seeing conditions (i.e., $\lesssim$ 1$^{\prime\prime}$) for the reduced data and provides a representative estimate of the effective PSF across our data. We will return to re-examine this assumption later (\S\ref{sec:phot_cal}).

To characterize the ILB, we also rely on measurements derived from the Disco Ball dataset. We construct a narrow-band image centered on the redshifted [O{\sc III}]$\lambda5007$ line with a 5 \AA\ spectral window. This bandwidth corresponds to approximately 300 \kms, sufficient to capture the full extent of the emission line while accounting for the galaxy's possible rotation and velocity dispersion. We identify the three brightest emission-line clumps in the Disco Ball field and extract their spectral profiles. From these, we measure an average FWHM of 2.37 $\pm$ 0.05 \AA\ in the spectral direction, corresponding to $\sigma = 1.0$ \AA\ or 68 \kms. Because this width is significantly larger than the intrinsic velocity dispersion expected for stellar clusters, we attribute the measured broadening entirely to instrumental effects.
We adopt this value as representative of the ILB and apply it uniformly to all targets in our sample. This approach assumes that the instrumental configuration is stable across observations.

\begin{figure*}
    \centering
	\includegraphics[width=0.85\linewidth]{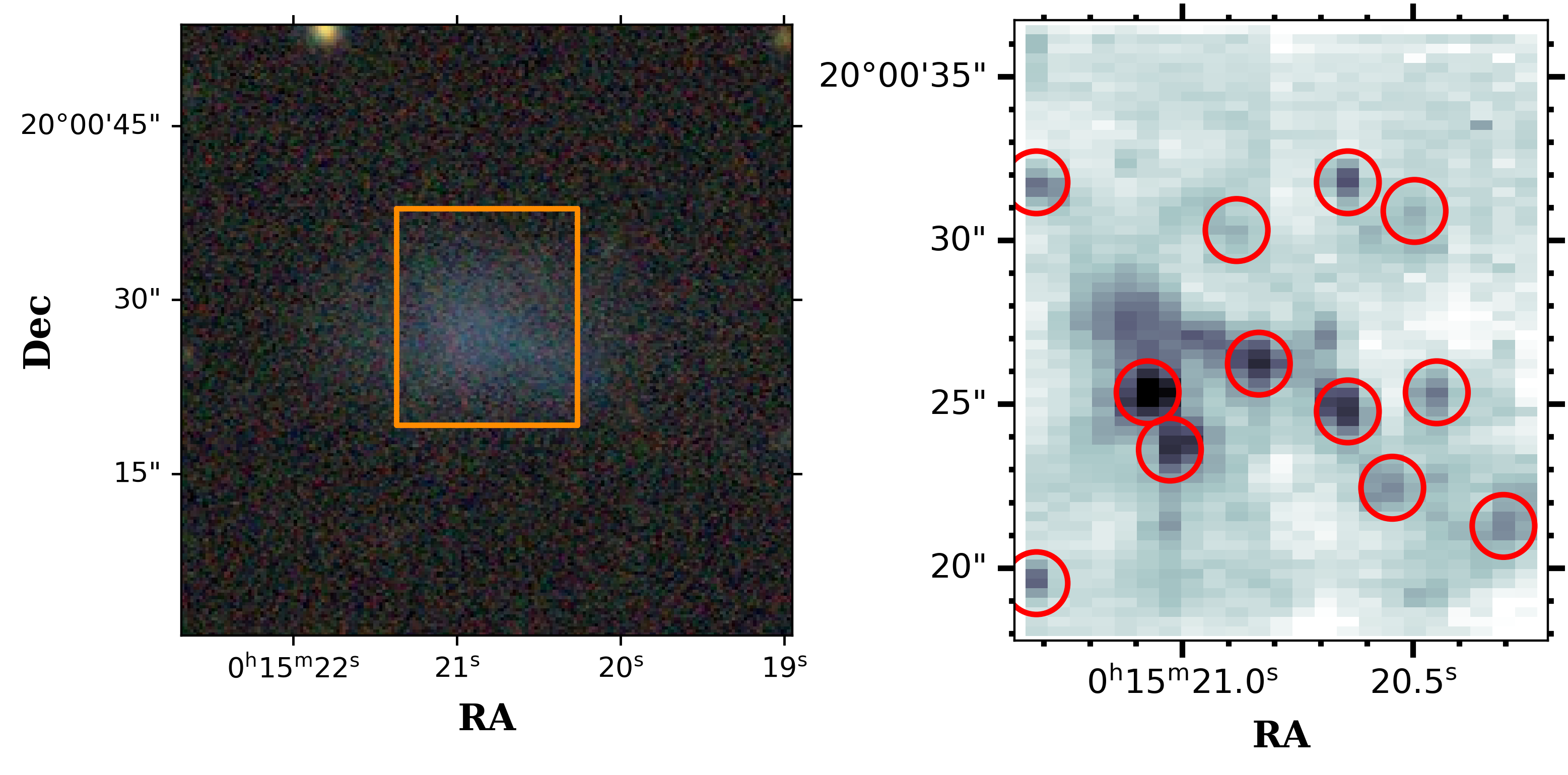}
    \caption{\textit{Left}: Legacy Surveys image of one example galaxy, SMDG0015208+200027, obtained from the the Legacy Surveys online viewer (\url{ https://www.legacysurvey.org/viewer}). The orange box indicates the field of view of our KCWI data. \textit{Right}: Cluster detection on the background-subtracted KCWI broadband image of this galaxy, where the broadband image is constructed by collapsing the full KCWI spectral range. Red circles indicate identified clusters, defined as groups of at least two adjacent pixels with pixel values exceeding $2\sigma$ above the local background. This comparison highlights some of the advantages provided by the KCWI data and how an apparently smooth LSB galaxy contains significant structure.
    }
    \label{fig:cluster_detection}
\end{figure*}

\subsection{Cluster Finding}
\label{sec:finding}

In this subsection, we outline the procedure used to identify compact luminous clumps within galaxies. Throughout this work, we refer to these clumps as stellar clusters, although their physical nature is not fully constrained. In particular, we do not directly establish whether these sources are gravitationally bound systems.
Nevertheless, given the absence of alternative classes of objects that exhibit similar compact stellar overdensities in UDGs, we adopt the working assumption that these sources are structurally analogous to star clusters identified in other galaxies.

To search for compact sources, we first reconstruct the KCWI data cube into two-dimensional images (hereafter ``slices'') that isolate different spectral components. We construct narrow-band emission line slices centered on the three prominent emission lines used in this analysis: [O{\sc II}], H$\beta$, and [O{\sc III}]. We also construct a broadband slice from the full spectral range, including all emission-line contributions and the stellar continuum.

The source-detection step uses the \texttt{Source Extractor Python library} \citep[SEP;][]{sep}, a Python implementation of the original \texttt{Source Extractor} algorithm \citep{bertin}. We apply the procedure described below independently to each slice.

\subsubsection{Clusters in the broadband image}
\label{sec:finding_cc}

We begin by identifying clusters in the broadband slice, constructed from the full spectral range, for all galaxies in the sample. Because these sources are embedded within the diffuse stellar light of their host galaxies, spatial variation in the background can significantly affect their detectability. 
To correct for this background, we use the SEP package to model and subtract a spatially varying background while also estimating the spatially varying background noise. The noise estimates are used later in our source detection. We adopt a background mesh size of 2.04 $\times$ 2.04$^{\prime\prime}$ (3$\times$7 spaxels), which provides a balance between capturing large-scale variations in the diffuse light and avoiding overfitting small-scale structure. The scale corresponds approximately to 2--3 times the PSF FWHM and is therefore sufficient to trace the underlying background without suppressing compact sources. 

After subtracting the background, we use SEP to detect sources. We classify a detection as a cluster when it consists of at least two continuous spaxels with flux values exceeding $2\sigma$ above the local background. Figure~\ref{fig:cluster_detection} presents an example of the resulting detection map, with the identified clusters highlighted. We apply this procedure uniformly to all targets to ensure consistency across the sample.

For each detected cluster, we return to the image prior to background subtraction and measure an instrumental aperture flux by summing the spaxel values within a circular aperture of radius equal to the PSF FWHM. To avoid contamination from neighboring sources, we first mask all other identified clusters using apertures with radii equal to the PSF FWHM. 
We estimate and subtract the local background using an annular region with inner and outer radii of 1.5 and 3.5 times the PSF FWHM, respectively.  We do not apply an aperture correction, but we also do not apply an aperture correction to our calibration sources when converting to apparent magnitudes (\S\ref{sec:phot_cal}). 

Background galaxies can contaminate the cluster catalog, and color information provides a useful way to identify obvious red background systems. Because the KCWI wavelength range alone does not provide a sufficiently red bandpass for this purpose, we use external imaging from the Legacy Survey \citep{dey}. We measure the $g-z$ colors from the Legacy Survey images using aperture photometry with a 5-pixel-radius aperture ($\sim$ 1.3$^{\prime\prime}$) centered on each source. \cite{Peng2006} showed that for host galaxies with integrated colors of $g-z \lesssim 1.3$ there are few if any GCs with $g-z \geq 1.5$. Because the median integrated color of our host-galaxy sample is $g-z = 0.75$, we expect most GCs associated with these galaxies to be bluer than this limit. We therefore remove sources that are robustly redder than $g-z = 1.5$, classifying those with $(g-z) - \sigma_{g-z} > 1.5$ as likely background galaxies.

\subsubsection{Clusters in the emission slices}
\label{sec:finding_ec}

We next search for clusters in the narrow emission-line slices, which trace regions of ongoing or recent star formation (age $\lesssim$ 20 Myr). Although such sources are not typically categorized as GCs, their detection provides valuable information on recent star formation activity, ionization conditions, and kinematics within the galaxies.

To isolate these features, we construct narrow-band slices centered on key emission lines: [O{\sc II}] $\lambda3727$, H$\beta$, and [O III] $\lambda5007$. For each line, we use a 5 \AA\ bandpass, corresponding to approximately twice the FWHM of the ILB, centered at the redshifted wavelength of the corresponding transition. We subtract the local continuum using adjacent wavelength regions offset by $\pm 5$ \AA\ from the line bandpass.

\begin{figure}
	\includegraphics[width=\columnwidth]{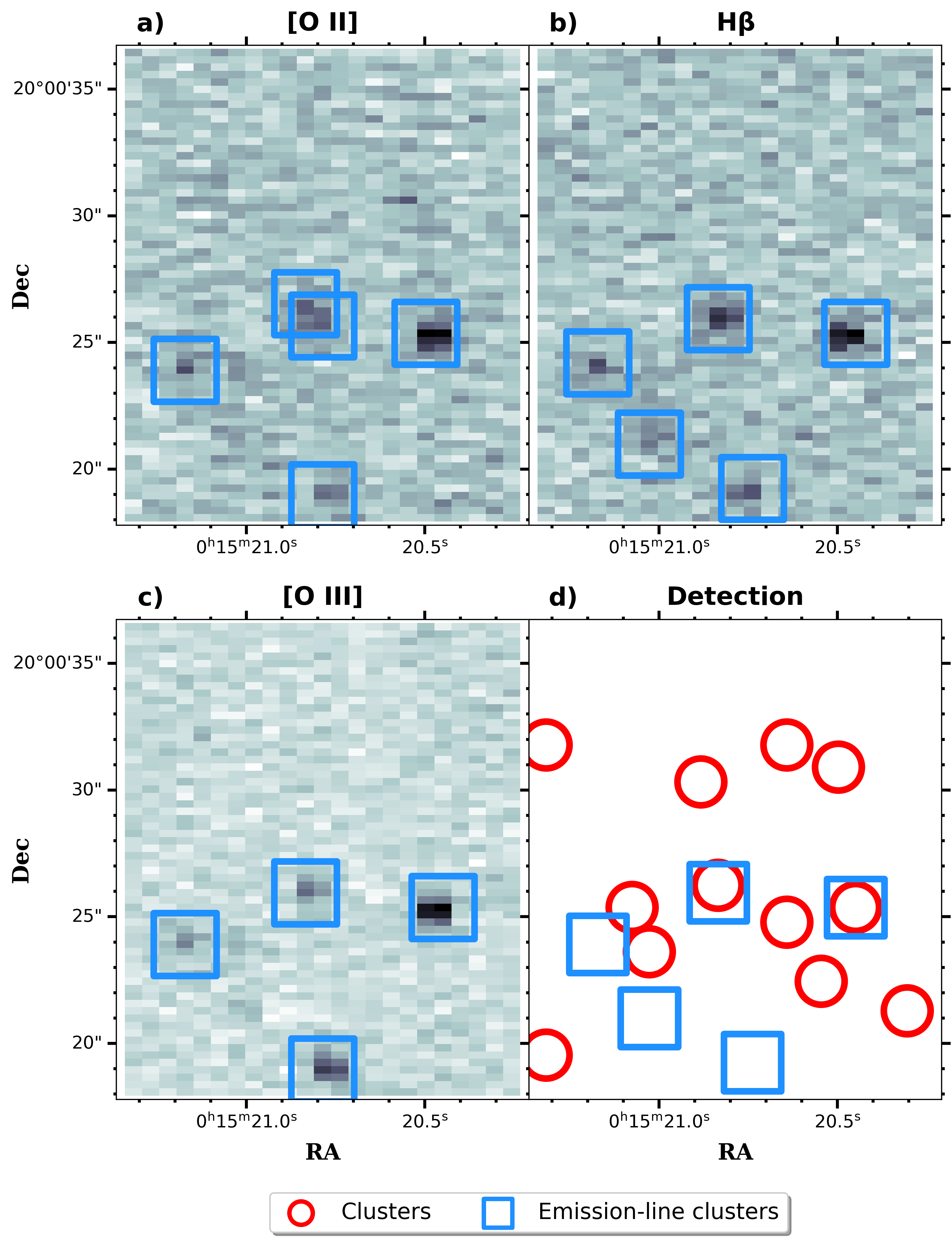}
    \caption{Cluster detection in the emission slices of SMDG0015208+200027 (same galaxy as in Figure~\ref{fig:cluster_detection}). Panels (a)--(c) show the identified clusters in the [O{\sc II}]$\lambda3727$, H$\beta$, and[O{\sc III}]$\lambda5007$ slices, respectively. Blue squares mark emission-line clusters, identified as $\ge 2$ contiguous pixels exceeding $3\sigma$ above the background.   
    Panel (d) summarizes all detected sources. Clusters detected in the broadband image are shown as red circles (Figure~\ref{fig:cluster_detection}), while those detected in the emission-line slices are shown as blue squares. Some clusters are detected in both the full spectral range image and in emission-line slices, while others have no emission lines or are detected only as emission-line sources.
    }
    \label{fig:EC_detection}
\end{figure}

We identify emission-line clusters using the procedure described in \S\ref{sec:finding_cc}, but with a more stringent spaxel value threshold of $> 3\sigma$ above the local background to ensure robust identification of emission-line features. We perform the detection independently in the three slices and classify a source as an emission-line cluster if it is detected in at least one of these slices. We match detections across different emission lines by position, so sources detected in multiple slices are counted only once. Additionally, we merged detections separated by less than one PSF FWHM and treated them as a single source because such close pairs cannot be robustly separated at the spatial resolution of the KCWI data.
Figure~\ref{fig:EC_detection} shows an example of the resulting detections, with emission-line clusters highlighted in the individual slices and summarized alongside the broadband-selected clusters. We apply this methodology consistently across all galaxies and all emission-line slices.

\subsection{Photometric Calibration}
\label{sec:phot_cal}

To determine whether the detected compact sources occupy the luminosity range expected for GCs, we require a photometric calibration for the KCWI detections. Although we use the Legacy Survey images to reject obvious background galaxies, we cannot rely on those images for the photometry of the compact sources because most of our detections are too faint to be measured reliably in the Legacy imaging. We therefore convert the KCWI fluxes to a standard $g$-band magnitude scale. 

We perform synthetic photometry on each extracted spectrum with the \texttt{PYPHOT} package \citep{zenodopyphot}, integrating each spectrum through the SDSS $g$-band transmission curve. We then convert the resulting instrumental fluxes to apparent $g$-band magnitudes by normalizing them to the Disco Ball NSC, for which a previously measured $g$-band magnitude is available \citep{khim+24_nsc}. Because the KCWI spectra are not spectrophotometrically calibrated, this approach assumes KCWI has uniform sensitivity across the $g$-band passband. This approximation is reasonable because the BL grating throughput varies relatively smoothly across most of the $g$-band \citep{kcwi}.

\begin{figure}
	\includegraphics[width=\columnwidth]{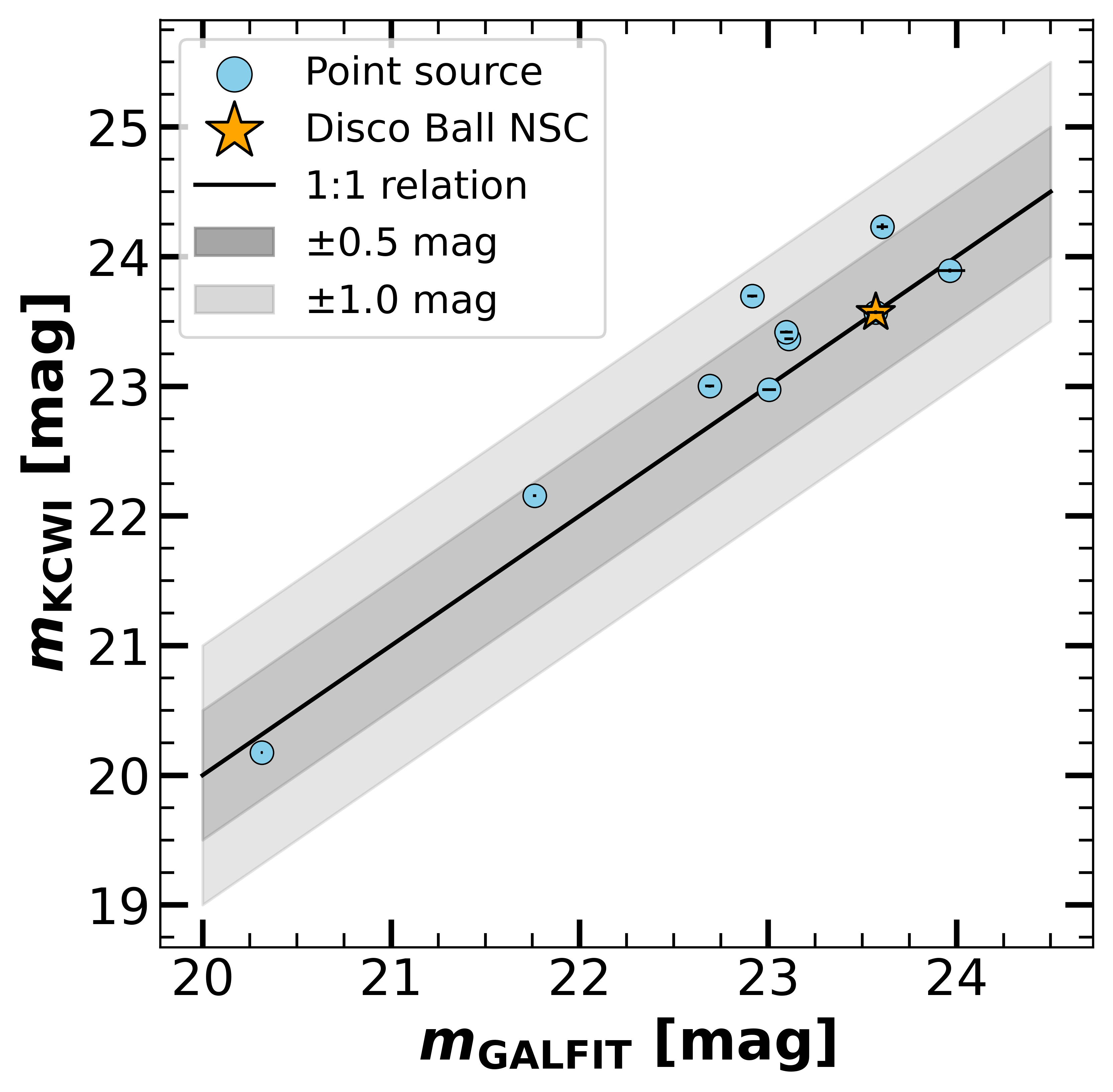}
    \caption{Comparison between KCWI and GALFIT $g$-band magnitudes. For point sources located within the KCWI field of view, we measure KCWI $g$-band magnitudes from the extracted spectra and photometric $g$-band magnitudes using GALFIT on the Legacy Survey images. Blue circles represent these point sources, and the orange star marks the Disco Ball NSC, which we use as the reference source for the KCWI flux calibration. The black solid line shows the one-to-one relation, while the dark and light gray shaded regions show offsets of $\pm0.5$ and $\pm1.0$ mags, respectively. The measurement uncertainties are smaller than the plotted symbols.
    }
    \label{fig:phot}
\end{figure}

To check the quality of the calibrated photometry, we identify bright point sources within the KCWI fields that are visible in the Legacy Survey viewer images. We fit the $g$-band imaging data with \texttt{GALFIT} \citep{Peng_2010} using a setup similar to that used in \cite{khim+24_nsc}. To be specific, we model the galaxy's diffuse light with a S\'ersic profile and each compact source with a PSF component. We use the PSF model and the inverse-variance maps provided by the Legacy Survey. These fits provide independent magnitudes for the point sources, measured separately from the KCWI spectra. We then compare these GALFIT-based magnitudes with KCWI-based magnitudes derived using the Disco Ball NSC calibration, as shown in Figure~\ref{fig:phot}. The blue circles mark the point sources used for the test, and the orange star marks the Disco Ball NSC, which defines the KCWI magnitude zero point. 

This comparison shows that the magnitudes derived for some sources agree well, including for the brightest source, but that there is a tendency for the KCWI magnitudes to be fainter than the Legacy ones. The mean difference is 0.24 mag.
One possibility is that our assumption of a uniform PSF across the images biases our result. To test this possibility, we compare the photometric differences in Figure~\ref{fig:phot} with the FWHM values reported in the image headers. We find: 1) no large differences in FWHM within the sample, 2) that the Disco Ball images have a FWHM that is near the middle of the range, and 3) no trend between the photometric differences and the FWHM. We conclude that PSF variations are not responsible for the observed photometric differences. Instead, 
we attribute the photometric differences to differences in how the background is estimated in the two datasets. In KCWI, we measure a local value, while in Legacy, we adopt a uniform background and model the galaxy light.
Because we have slightly higher confidence in the KCWI magnitudes because the background is complex and a local background is probably best, we conclude that our magnitude uncertainties are $\lesssim$ 0.29 mag (rms) and that if anything, the KCWI magnitudes are slight ($\lesssim 0.24$ mag) underestimates.

\subsection{Spectral Analysis}
\label{sec:spec}

In this subsection, we describe how we extract spectra from the KCWI data to investigate the kinematic properties of different galactic components across the full sample. For the diffuse component, we combine spaxels within 0.5 \re\ of each galaxy center after masking the identified clusters using circular apertures with a radius of one PSF FWHM. 
For each cluster, we extract a spectrum by summing the spaxels within a circular aperture with a radius of one PSF FWHM centered on the cluster. We then subtract the local background using a local background estimated from the same annular region used for the flux measurements (\S\ref{sec:finding_cc}).

We derive stellar and gas kinematics using the \texttt{Penalized PiXel-Fitting} package \citep[pPXF;][]{ppxf}. The fitting procedure simultaneously models three spectral components: a stellar continuum, Balmer emission lines, and oxygen emission lines. 
We model the stellar component using the \texttt{X-shooter Spectral Library} \citep[XSL;][]{xsl}. To account for residual mismatches in the continuum shape between the data and templates, we allow pPXF to apply a fourth-order additive Legendre polynomial in the fit. For instrumental spectral broadening, we use the ILB value derived in \S\ref{sec:quality}. We initialize the velocity using the redshifts reported by \cite{loraine} and the velocity dispersion at 200 \kms, approximately three times the ILB, following the recommendation of the pPXF documentation.

We apply this fitting procedure to the extracted spectra as described above, the diffuse galaxy component (within 0.5 \re\ and excluding cluster regions) and all identified clusters.
To ensure reliable kinematic measurements, we reject fits when the peak amplitude of the best-fit model does not exceed the continuum noise level or when the uncertainty in the recessional velocity is greater than 30 \kms.

\section{Results}
\label{sec:results}

\subsection{Sample Galaxies}
\label{sec:gal}

We first compare our redshift measurements with those from \cite{loraine} for galaxies with measurements in both studies. The two sets of measurements agree well, with a mean fractional difference of 1.0\%. Two of the 44 galaxies lacked redshift measurements in \cite{loraine}. For one of these galaxies, SMDG0000334+165424, we measure a redshift of $z=0.00290\pm0.00002$. For the other, SMDG2246485-105427, we also do not obtain a reliable redshift from the current data.

\begin{figure}
	\includegraphics[width=\columnwidth]{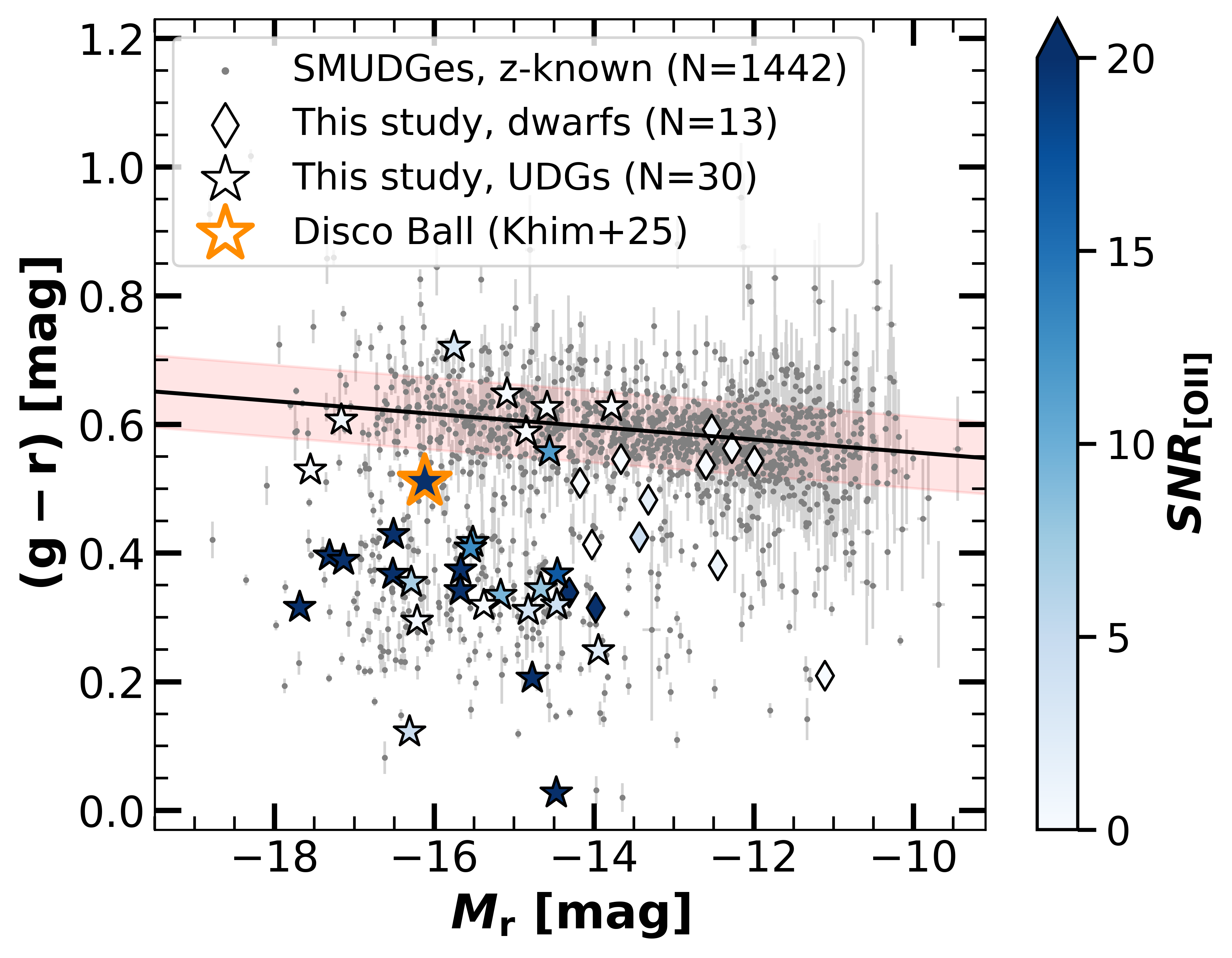}
    \caption{Color-magnitude diagram of the galaxies in our sample. We show the 43 KCWI target galaxies selected from the SMUDGes catalog by \cite{loraine}, using stars for UDGs (\re $\geq 1.5$ kpc) and diamonds for the remaining dwarfs. Symbol colors indicate the SNR of [O{\sc II}] emission measured from the galaxy spectrum after masking the identified clusters, providing a tracer of recent or ongoing star formation. We highlight the Disco Ball with an orange outline \citep{Khim+25}. Gray points show SMUDGes galaxies with spectroscopic or estimated redshifts \citep{smudges5}. The solid black line and shaded region represent the fitted red sequence and its scatter, respectively \citep{khim+24_nsc}. 
    }
    \label{fig:CMD_OII}
\end{figure}

We present the color-magnitude diagram of our sample in Figure~\ref{fig:CMD_OII}, along with that of the SMUDGes galaxies with spectroscopic or estimated redshifts. We adopt the photometric parameters from the SMUDGes catalog for this plot. The KCWI targets span the range of galaxy properties, although they overrepresent the bluer colors relative to the full sample, which suggests ongoing or recent star formation in a large fraction of the KCWI sample. We also show the fitted red sequence and its scatter \citep{khim+24_nsc}. 

Using these redshift measurements, we also evaluate whether the sample galaxies satisfy the UDG size criterion. Among the 43 KCWI galaxies with reliable redshifts, 30 satisfy the UDG size criterion of \re $\geq 1.5$ kpc. 
The UDGs are not confined to the red sequence, consistent with the known diversity of field UDGs \citep[e.g.,][]{Prole+19_field}. The Disco Ball lies in the green-valley region, as previously discussed in \cite{Khim+25}.

The symbol's color in Figure~\ref{fig:CMD_OII} corresponds to the signal-to-noise ratio of the [O{\sc II}] emission measured from the galaxy spectrum, after masking the identified cluster regions. Because [O{\sc II}] emission is sensitive to recent or ongoing star formation, this measurement provides a qualitative tracer of star-forming activity in each galaxy outside of the identified emission line clusters. Galaxies near the red sequence generally show weak or undetected [O{\sc II}] emission, consistent with a lower level of recent star formation. In contrast, many galaxies in the blue cloud show strong [O{\sc II}] emission, supporting the interpretation that their blue colors reflect recent or ongoing global star formation. We find a similar trend when using [O{\sc III}] emission instead of [O{\sc II}]. 

All but three of the blue cloud galaxies satisfy our UDG size criterion. This result is broadly consistent with the established picture that field UDGs are often blue and star-forming, in contrast to the largely quenched UDG populations in dense environments. Our analysis extends this picture by using spatially resolved spectroscopy to identify recent or ongoing star formation within blue UDGs. 
We also identify rare cases of red-sequence UDGs that host emission lines (SMDG0045533-093353 and SMDG2337078+001240). This suggests that some may still retain localized or residual star formation activity.

\subsection{Cluster Populations}
\label{sec:NClu}

We now examine the cluster populations identified in \S\ref{sec:finding} and define the categories used throughout the following analysis. 
We classify the clusters detected in the emission-line slices as emission-line clusters (EC), regardless of whether they have continuum counterparts (\S\ref{sec:EC}). We define continuum clusters (CCs) as clusters detected in the broadband image but not in any emission-line slice. Separately, we identify NSC candidates from the broadband-detected sample based on their central locations and relative luminosities (\S\ref{sec:NSC}).

CCs are the closest analogs to GCs, although some may instead be intermediate-age stellar clusters rather than canonical old GCs. Questions about whether the term GC should exclusively refer to old, rich stellar clusters date back to the earliest studies of the compact stellar clusters in the Magellanic Clouds \cite[c.f.,][]{gascoigne,hodge_1961}. Without age measurements, we use the CC sample as the parent sample from which we later define a more restricted set of GCs (\S\ref{sec:GC}).
As \cite{Khim+25} noted, the inclusion of younger, brighter clusters in a ``GC sample" can lead to an overestimation of the size of the GC population or an underestimation of the distance if one assumes a standard GC luminosity function. We will return to this issue in \S\ref{sec:GCLF}. 

We summarize the number of detected clusters in each category for every galaxy in Table~\ref{table:cluster_numbers}. We measure each population within the largest galaxy-shaped elliptical aperture that lies fully inside the KCWI field of view, adopting each galaxy's photometric axis ratio and position angle. For the GC population, we additionally apply a radial correction to account for clusters outside the observed aperture and a photometric correction, based on the assumed GC luminosity function (GCLF), to account for clusters fainter than our detection limit (\S\ref{sec:detec_limit}).

\begin{table*}[t]
\centering
\caption{Number of Detected Clusters and Corrected Globular Clusters in the KCWI Fields}
\label{table:cluster_numbers}
\begin{tabular}{lcccc|cc}
\hline\hline
ID &
$R_{\rm Ell}/r_{\rm e}$ &
$N_{\rm EC,Ell}$ &
$N_{\rm CC,Ell}$ &
$N_{\rm GC,Ell}$ &
$N_{\rm GC,rad}$ & 
$N_{\rm GC,tot}$ \\
\hline
SMDG0000334+165424 & 0.89 & 0 & 9 & 8 & $19.1 \pm 6.8$ & $19.1 \pm 6.8$ \\
SMDG0015089$-$031837 & 0.84 & 3 & 2 & 2 & $4.2 \pm 3.0$ & \nodata \\
SMDG0015208+200027 & 0.81 & 3 & 5 & 5 & $11.9 \pm 5.3$ & \nodata \\
SMDG0033164+283311 & 0.94 & 2 & 4 & 4 & $6.0 \pm 3.0$ & \nodata \\
SMDG0037442+241228 & 0.67 & 2 & 2 & 2 & $11.2 \pm 8.0$ & \nodata \\
SMDG0038365$-$064207 & 0.49 & 3 & 8 & 8 & $30.8 \pm 10.9$ & $59.3 \pm 21.0$ \\
SMDG0045533$-$093353 & 1.09 & 0 & 7 & 7 & $10.9 \pm 4.1$ & \nodata \\
SMDG0102539+305357 & 1.02 & 4 & 6 & 6 & $10.8 \pm 4.4$ & \nodata \\
SMDG0109254$-$014543 & 0.75 & 0 & 6 & 6 & $17.9 \pm 7.3$ & \nodata \\
SMDG0122562+084025 & 0.88 & 0 & 7 & 7 & $19.5 \pm 7.4$ & $19.5 \pm 7.4$ \\
SMDG0124406$-$013812 & 1.05 & 0 & 4 & 4 & $7.0 \pm 3.5$ & $19.5 \pm 9.7$ \\
SMDG0127038$-$082400 & 0.92 & 3 & 6 & 6 & $10.2 \pm 4.1$ & \nodata \\
SMDG0134336+005425 & 2.02 & 2 & 10 & 10 & $6.9 \pm 2.2$ & \nodata \\
SMDG0136469+034323 & 0.65 & 0 & 6 & 6 & $19.3 \pm 7.9$ & \nodata \\
SMDG0145431$-$084143 & 0.84 & 6 & 7 & 7 & $19.0 \pm 7.2$ & \nodata \\
SMDG0147005$-$142556 & 0.97 & 0 & 5 & 5 & $8.6 \pm 3.9$ & $8.6 \pm 3.9$ \\
SMDG0148234$-$131444 & 1.02 & 0 & 2 & 1 & $2.3 \pm 2.3$ & $2.3 \pm 2.3$ \\
SMDG0152163$-$044242 & 0.29 & 5 & 3 & 3 & $26.2 \pm 15.1$ & \nodata \\
SMDG0224231$-$031059 & 0.58 & 0 & 6 & 6 & $20.4 \pm 8.3$ & $20.4 \pm 8.3$ \\
SMDG0321028$-$042741 & 0.93 & 7 & 3 & 3 & $7.8 \pm 4.5$ & \nodata \\
SMDG0455129$-$031122 & 1.07 & 0 & 9 & 9 & $14.0 \pm 4.7$ & $51.9 \pm 17.3$ \\
SMDG0459164$-$072243 & 0.91 & 0 & 5 & 4 & $11.1 \pm 5.6$ & \nodata \\
SMDG1756253+331055 & 0.67 & 0 & 5 & 5 & $19.2 \pm 8.6$ & \nodata \\
SMDG2104498+002509 & 1.27 & 2 & 1 & 1 & $1.3 \pm 1.3$ & \nodata \\
SMDG2113130+014217 & 1.43 & 5 & 9 & 8 & $9.5 \pm 3.4$ & \nodata \\
SMDG2207591$-$100713 & 1.02 & 5 & 7 & 7 & $12.9 \pm 4.9$ & \nodata \\
SMDG2217328$-$132522 & 0.53 & 1 & 7 & 7 & $27.4 \pm 10.3$ & \nodata \\
SMDG2227407$-$122104 & 0.87 & 4 & 2 & 2 & $3.3 \pm 2.3$ & \nodata \\
SMDG2235086+011040 & 1.21 & 0 & 4 & 2 & $3.9 \pm 2.8$ & \nodata \\
SMDG2235405+233627 & 0.92 & 0 & 5 & 5 & $9.7 \pm 4.3$ & \nodata \\
SMDG2239446+180218 & 0.86 & 5 & 0 & 0 & \nodata & \nodata \\
SMDG2240062+223715 & 1.21 & 0 & 10 & 10 & $19.3 \pm 6.1$ & \nodata \\
SMDG2241452+222153 & 1.01 & 5 & 3 & 3 & $4.8 \pm 2.8$ & \nodata \\
SMDG2246485$-$105427 & 0.73 & 0 & 10 & 0 & \nodata & \nodata \\
SMDG2251037$-$015247 & 1.37 & 0 & 6 & 6 & $7.9 \pm 3.2$ & $13.0 \pm 5.3$ \\
SMDG2300066+154949 & 1.28 & 0 & 7 & 7 & $9.7 \pm 3.7$ & $21.7 \pm 8.2$ \\
SMDG2302155+300423 & 1.17 & 0 & 12 & 9 & $16.6 \pm 5.5$ & \nodata \\
SMDG2323184+172116 & 1.10 & 0 & 2 & 2 & $4.1 \pm 2.9$ & \nodata \\
SMDG2329214$-$063209 & 0.96 & 5 & 3 & 3 & $4.8 \pm 2.8$ & \nodata \\
SMDG2337078+001240 & 0.79 & 1 & 11 & 11 & $28.7 \pm 8.7$ & $39.2 \pm 11.8$ \\
SMDG2338066+003203 & 0.98 & 5 & 5 & 5 & $9.3 \pm 4.1$ & \nodata \\
SMDG2343135$-$083901 & 0.81 & 3 & 11 & 10 & $26.6 \pm 8.4$ & \nodata \\
SMDG2359136+164655 & 1.10 & 0 & 5 & 3 & $7.1 \pm 4.1$ & $7.1 \pm 4.1$ \\
SMDG2359282+150125 & 0.48 & 2 & 5 & 5 & $27.0 \pm 12.1$ & \nodata \\
\hline
\end{tabular}

\vspace{0.5em}
\begin{minipage}{0.90\textwidth}
\footnotesize
\textit{Note.} $R_{\rm Ell}/r_{\rm e}$ represents the semi-major axis of the largest elliptical aperture fully enclosed within the observed field of view, normalized by the galaxy's effective radius. $N_{\rm EC,Ell}$, $N_{\rm CC,Ell}$, and $N_{\rm GC,Ell}$ show the number of emission-line, continuum, and globular clusters detected within this elliptical aperture, respectively. $N_{\rm GC,rad}$ is the GC count after applying the radial incompleteness correction (\S\ref{sec:GC}), and $N_{\rm GC,tot}$ is the total GC count after applying both the radial and photometric incompleteness corrections (\S\ref{sec:detec_limit}). The quoted uncertainties are Poisson uncertainties.
\end{minipage}
\end{table*}

\begin{figure}
	\includegraphics[width=\columnwidth]{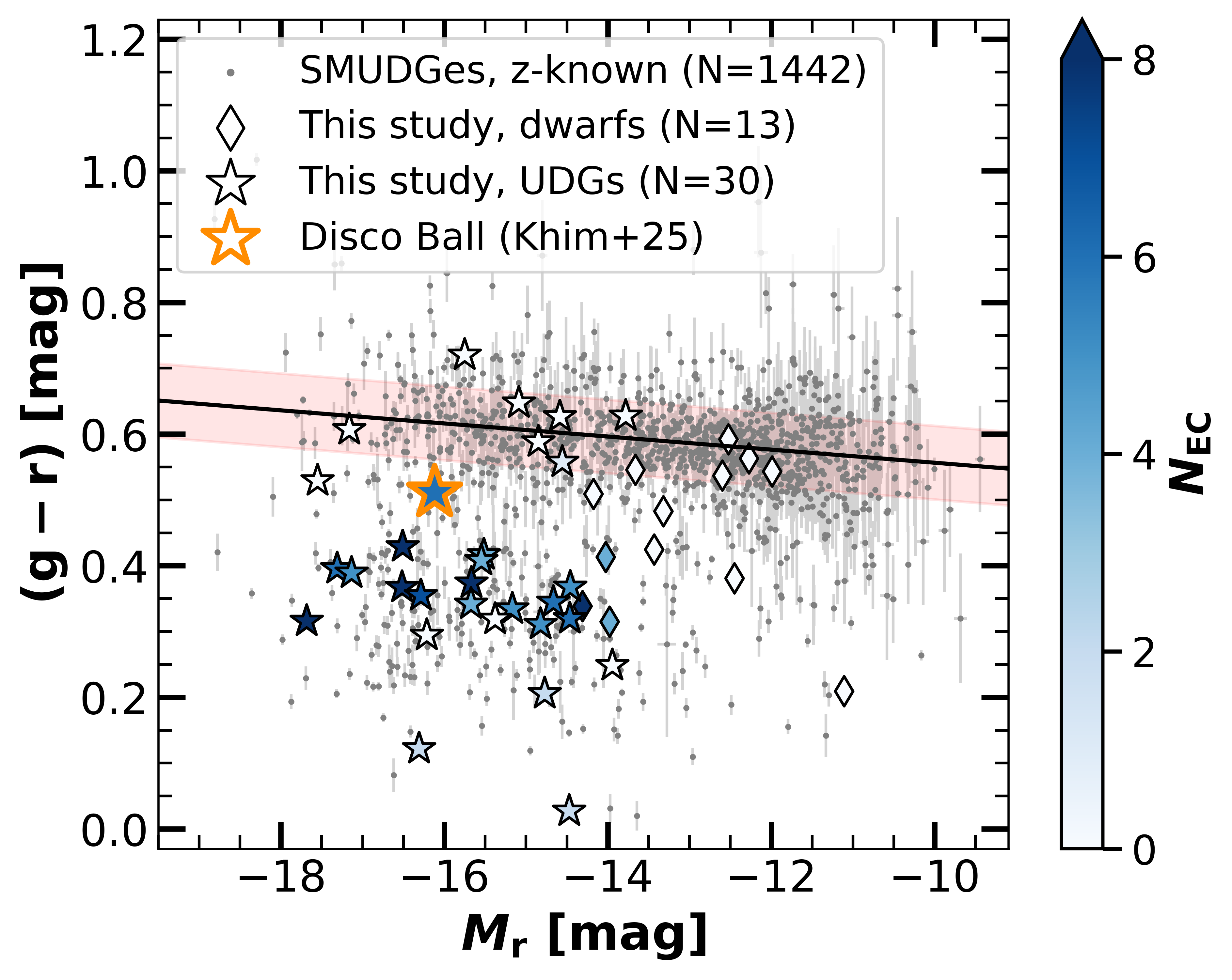}
    \caption{The number of identified ECs in the same format as Figure~\ref{fig:CMD_OII}. All galaxies near the red sequence host either zero or one EC, whereas blue-cloud galaxies typically host multiple ECs. The Disco Ball lies in the green valley and hosts multiple ECs. 
    }
    \label{fig:CMD_EC}
\end{figure}

\subsubsection{Emission-line clusters}
\label{sec:EC}

ECs have strong enough spectral features to allow individual redshift measurements with pPXF. We remove two sources whose redshifts differ clearly from that of the host galaxy ($\Delta v > 100$ \kms) and so have confidence that all of the ECs we present are associated with the host galaxy.

In Figure~\ref{fig:CMD_EC}, we present a color-magnitude diagram, where we color-code each target by the number of ECs detected. Among the 43 galaxies in our sample, we identify at least one EC in 24 systems. The distribution of ECs broadly mirrors the trend seen in Figure~\ref{fig:CMD_OII}. Galaxies near the red sequence generally show no detected ECs, with only one exception, and even that galaxy hosts only a single EC, indicating little current star formation.
In contrast, the blue cloud population almost always hosts multiple ECs. Only three blue cloud galaxies lack detected ECs. 

The emission-line spectra of the ECs confirm that these compact sources are associated with the host galaxies and trace very recent star formation. Thus, while the [O{\sc II}] and [O{\sc III}] measurements (Figure~\ref{fig:CMD_OII}) provide a galaxy-scale tracer of recent star formation, the ECs identify the compact regions where this activity is most clearly localized.

However, the correspondence between EC counts and emission-line strength is not one-to-one. Among the full sample, 17 galaxies have no detected ECs, but five of these still show [O{\sc II}] emission with SNR $>2$ in their cluster-masked spectra. This residual emission indicates that recent star formation can be present even when it does not appear as ECs in our data.
Some of the signal may come from residual flux from partially masked clusters, but it may also trace fainter or more diffuse star-forming regions that are not detected as compact ECs. The relative contributions of compact ECs and residual galaxy-scale emission therefore suggest that recent star formation varies not only in strength, but also in the degree to which it is concentrated into high-surface-brightness regions.

\begin{figure*}
	\includegraphics[width=\linewidth]{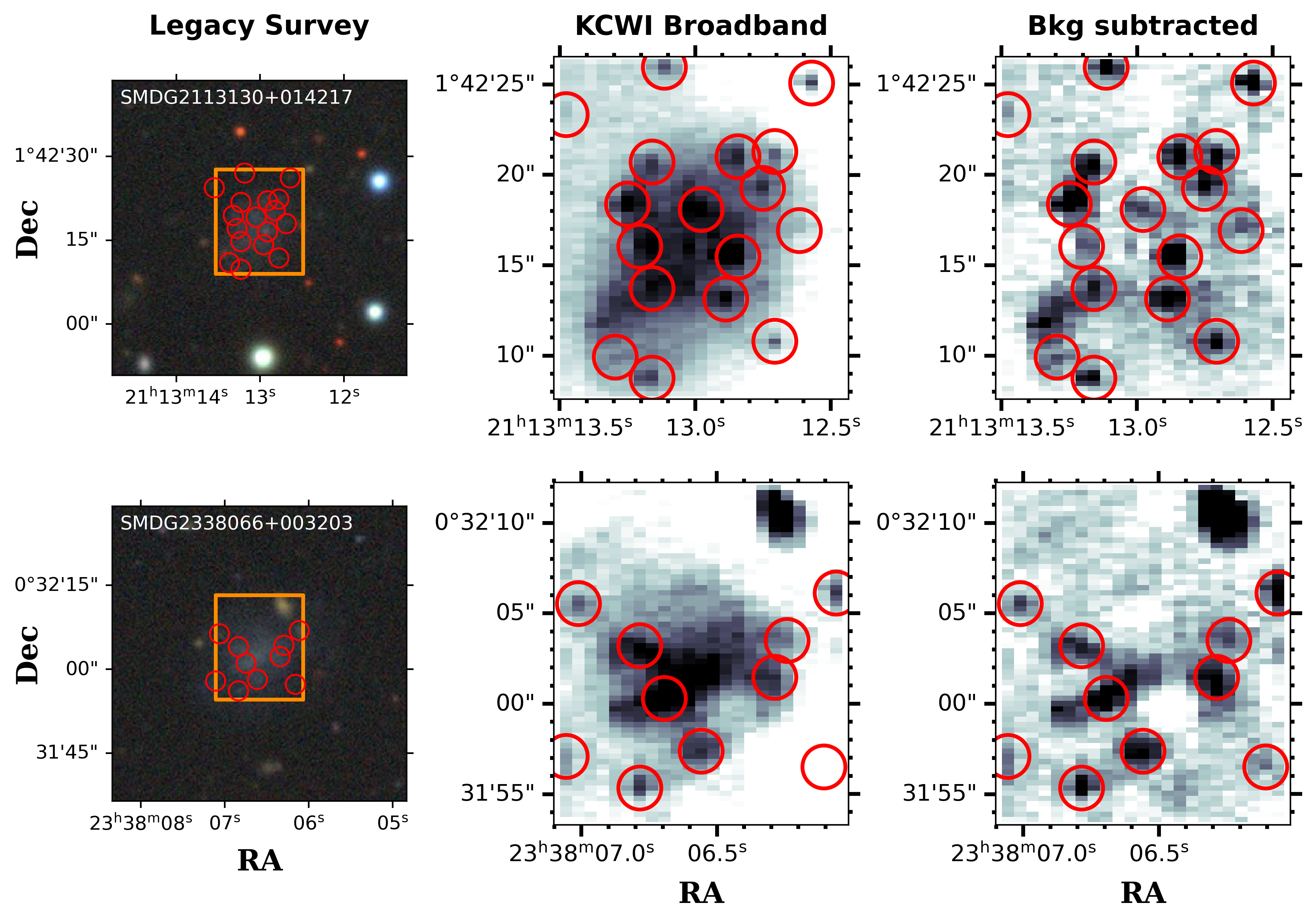}
    \caption{Example of GC detections in two sample blue galaxies. From left to right, each row shows the Legacy Survey color image, the KCWI broadband image, and the spatially varying background-subtracted image. Red circles present the identified GCs, and the orange boxes in the Legacy Survey images indicate the field of view of the KCWI data. Some detections may correspond to star clusters, whereas others may represent clumpy structure within the host galaxy.
    }
    \label{fig:GC_compare}
\end{figure*}

\subsubsection{Globular clusters}
\label{sec:GC}

As noted above, the interpretation of CCs (i.e., clusters undetected in the emission-line slices) is complicated by the unknown ages of the clusters, which introduce substantial uncertainties in the inferred stellar masses and long-term survival probabilities \citep[e.g.,][]{chandar,gieles_2011,webb_2024}. Neglecting this caveat for the moment, we construct our sample of GC candidates from the CC sample. We first remove sources with an absolute $g$-band magnitude outside the range expected for typical GCs, adopting $-12 < M_g < -4$ mag \citep{2009Jordan_GC_NSC_Lum}. 
Next, we fit the spectrum of each source to measure a redshift. We do not obtain reliable redshift measurements for most of these sources, but when we do, we reject objects identified as clear foreground stars at $z\approx0$ or background galaxies at $z>0.1$. We reject a total of seven objects on this basis.
We also test whether structural parameters, such as concentration index and ellipticity, can distinguish clusters from background galaxies. These quantities provide no clear separation, likely because of the limited KCWI spatial resolution and underlying galaxy structure.

Although we refer to these objects as GCs, this terminology should be interpreted with caution. We cannot confirm that every individual source is an actual GC, primarily because reliable redshift measurements are unavailable for many of the detected clusters, except for a few bright clusters, and also because our ground-based imaging is insufficient to confirm the compact nature of the sources. Figure~\ref{fig:GC_compare} highlights this ambiguity with two blue galaxies that have similar $g-r$ colors ($g-r=0.43$ and $0.45$, respectively). In the first galaxy (SMDG2113130+014217), the broadband image reveals several bright, compact sources with angular sizes comparable to the PSF. These sources are therefore unresolved in our data and are plausible compact stellar clusters. However, in the second galaxy (SMDG2338066+003203) some identified sources may instead trace clumpy substructure within the galaxy itself. 
Follow-up integral field spectroscopy with higher angular resolution or deep, high-resolution imaging will allow us to distinguish between these possibilities and better constrain the cluster population. These factors limit our ability to reach conclusions regarding the nature of the GC populations in these galaxies, other than to confirm their general presence and highlight galaxies with potentially rich or poor GC systems.

\subsubsection{Nuclear star clusters}
\label{sec:NSC}

NSCs are compact, luminous stellar systems located near the centers of galaxies across a wide range of morphologies, environments, and stellar masses \citep[e.g.,][]{Lauer2005,Georgiev2014,Baldassare2014,sanchez2019b}. They are typically only a few to several tens of parsecs in size and are generally brighter than ordinary GCs \citep{Cote2006,neumayer}. 

We identify NSC candidates among the clusters detected in the broadband image, regardless of whether they are also detected in the emission-line slices. Following \cite{lambert} and \cite{khim+24_nsc}, we adopt a fiducial search radius of 0.1\re, motivated by the observed statistical excess of compact sources near galaxy centers. We further require the candidate to be at least 0.5 mag brighter than any other detected clusters within \re. We classify sources that satisfy both criteria as NSC candidates.

We identify NSC candidates in 3 out of 44 galaxies: Disco Ball (SMDG0038365-064207), SMDG0015089-031837, and SMDG2251037-015247. The latter two candidates were not included in \cite{khim+24_nsc}. The candidate in SMDG0015089-031837 is not visible in the shallower Legacy Survey imaging. The candidate in SMDG2251037-015247 is visible in the Legacy Survey data, but its host galaxy is not in the \cite{smudges5} catalog because it failed one of the additional selection cuts applied between the times when the KCWI SMUDGes sample was selected and when we converged on the final criteria applied by \cite{smudges5}. 

We spectroscopically confirm that all three NSC candidates are associated with their host galaxies. Their absolute $g$-band magnitudes are $M_g = -10.34\pm0.04$, $-9.77\pm0.02$, and $-8.64\pm0.02$ mag for SMDG0015089-031837, the Disco Ball, and SMDG2251037-015247, respectively. These values fall within the luminosity range expected for NSCs \citep[e.g.,][]{Cote2006, khim+24_nsc}. Assuming the same stellar mass-to-light ratio for each NSC and its host, their NSC-to-host stellar mass ratios are $(3.4-9.6)\times10^{-3}$, and all three lie within the scatter of the NSC mass--host stellar mass relation reported by \cite{khim+24_nsc}.

The brightest candidate, in SMDG0015089-031837, shows narrow [O{\sc II}], H$\beta$, and [O{\sc III}] emission lines. The galaxy-light-subtracted NSC spectrum contains significant Balmer and [O{\sc III}] emission, with [O{\sc III}]/H$\beta$ $=2.7\pm0.7$ and [O{\sc III}]/[O{\sc II}] $=2.7\pm0.7$. The high-ionization line ratios open the possibility of weak nuclear activity in this NSC. However, we do not detect He{\sc II} or [Ne{\sc III}] at high significance, and the KCWI wavelength coverage does not include the red optical lines needed for standard BPT diagnostics \citep{BPT}. We therefore cannot distinguish weak AGN activity from compact low-metallicity star formation using the current data alone. We detect no emission lines in the other two candidates.

The three host galaxies span a range of global colors and cluster populations, although SMDG2251037-015247 does not satisfy the UDG size criterion. SMDG0015089-031837 lies near the red edge of the blue cloud and hosts five emission-line clusters (one of which is the NSC candidate) and four GCs in the KCWI field of view. In contrast, SMDG2251037-015247 lies near the red sequence and hosts no emission-line clusters, but hosts eight GCs. The Disco Ball falls between these two systems, lying in the green valley, and hosts both a substantial population of emission-line clusters and GCs, with five and 13 objects, respectively. The different emission-line properties of the three candidates may reflect differences in their recent nuclear star formation or activity, although the small sample prevents a broader conclusion.

The three detections imply an NSC occupation fraction of $6.8\pm3.9\%$, where we estimate the uncertainty from Poisson statistics. This value is broadly consistent with the NSC occupation fraction found by \cite{khim+24_nsc} for the SMUDGes sample using the same normalized radial cut, $4.2\pm0.3\%$.

The diffuse and often clumpy morphologies of our galaxies, particularly the bluer systems, introduce uncertainty in the adopted galaxy centers. We assume that the center of each KCWI image coincides with the galaxy center, but pointing offsets may cause the two positions to differ. The 0.1\re\ criterion may therefore be too restrictive. As a check, we repeat the selection using a larger search radius of 0.2\re. This relaxed criterion identifies six additional NSC candidates. 
We confirm that NSC candidates in three host galaxies  (SMDG0045533-093353, SMDG0147005-142556, and SMDG2323184+172116) have redshifts consistent with those of their hosts. For the remaining three candidates (SMDG0124406-013812, SMDG0152163-044242, and SMDG2235086+011040), the spectra have insufficient SNR to determine whether they are associated with the host galaxies (i.e., velocity uncertainty greater than 30 \kms). If we assume that all six additional candidates are associated with their host galaxies and adopt the same stellar mass-to-light ratio for each NSC and its host, we find that all six lie within the scatter of the NSC mass--host stellar mass relation reported by \cite{khim+24_nsc}.

Including all six additional candidates increases the inferred NSC fraction to $20.5\pm6.8\%$, a factor of three above the value obtained with the 0.1\re\ cut. This relative increase is greater than that found by \cite{lambert}, who found that relaxing the radial cut from 0.1\re\ to 0.2\re\ increased the number of candidates by a factor of $\sim$1.4. This difference in results may reflect larger uncertainties in the adopted galaxy centers. However, given the small number of candidates and the sensitivity of the result to the adopted radial criterion, we do not draw a strong conclusion from this difference.

\subsection{Galaxy Kinematics}
\label{sec:kinematics}

In this section, we investigate the kinematic properties of galaxies in our sample. In our previous work on the Disco Ball \citep{Khim+25}, we found a rotating UDG hosting multiple ECs and detected consistent velocity gradients from the ECs and the diffuse stellar light. Motivated by this result, and by recent IFU studies showing that some UDGs host measurable stellar rotation \citep{LEWIS}, we examine whether similar systems are present in the current sample and assess how common such kinematics are among field UDGs.

The generally spheroidal morphologies of cluster UDGs suggest that many are pressure-supported rather than rotationally supported. 
Measurements of internal kinematics done so far \citep[e.g.,][]{vdk19,forbes+21} support this view, with typically no evident rotation and relatively large measurements of the stellar velocity dispersion. However, observations of H{\small I}-rich UDGs in the field show rotation \citep[e.g.,][]{leisman,pina+19,Pina+2020,karunakaran}, as did our investigation of the Disco Ball. Because our sample includes galaxies on both the red sequence and the blue cloud, we expect to have both dispersion- and rotation-supported galaxies in the sample.

Our spectral setup limits our ability to measure stellar velocity dispersions, which are typically 10--30 \kms \citep[e.g.,][]{forbes+21}, because the ILB exceeds 60 \kms. We therefore focus on spatially resolved velocity gradients. For galaxies with multiple ECs, we use their precisely measured velocities to trace the kinematics of the recently formed component, which likely follows the ordered kinematics of the gas disk. Where the diffuse stellar light has sufficient SNR, we also spatially bin the spectra and measure the bulk stellar velocity gradients.

\subsubsection{Rotation determined from ECs}
\label{sec:rot_ec}

Individual GCs have frequently been used as discrete kinematic tracers of galaxy kinematics. We extend this approach to ECs, using their positions and recessional velocities. ECs offer a practical advantage in LSB galaxies because their strong nebular emission lines yield precise velocities even when the diffuse stellar continuum does not provide sufficient SNR for a resolved velocity field. A small number of ECs distributed across a galaxy can therefore provide an efficient diagnostic of an ordered velocity gradient without requiring high-SNR absorption-line spectra over the full field of view. 

To estimate the rotation velocity of galaxies hosting ECs, we assume a linearly rising rotation curve and model the EC recessional velocities as a function of projected position along the photometric major axis. 
For each galaxy, we measure the velocity of each EC from its Balmer and oxygen lines.
We include only ECs with velocity uncertainties below 10 \kms\ and require at least three ECs per galaxy. For this analysis, unlike the EC counts reported in Table~\ref{table:cluster_numbers}, we use all ECs detected within the full KCWI field of view rather than only those within the elliptical aperture. After this cut, 19 galaxies remain. We then fit a simple linear velocity gradient using ordinary least squares and evaluate the fitted rotation velocity at 0.5\re.

We test the robustness of the fitting method by repeating the fits with weighted least squares after adding a 10 \kms\ uncertainty in quadrature to the formal velocity uncertainty of each EC. This term acts as a velocity-uncertainty floor, preventing points with very small formal uncertainties from dominating the fit and allowing for additional scatter from local gas motions or line-to-line velocity differences \citep{tamburro}. The resulting velocities are nearly identical to the ordinary least-squares results, but their uncertainties are roughly twice as large.

The EC-based approach recovers only the line-of-sight component of the rotation and, therefore, underestimates the intrinsic rotation velocity. We correct for this projection effect by assuming a thin-disk geometry.
We adopt the axis ratio from our photometric studies \citep{smudges2,smudges5}, and derive the corresponding inclinations under this assumption, following the procedure described in \cite{Khim+25}. To avoid highly uncertain corrections for nearly face-on systems, we exclude the two galaxies with inferred inclination below $30^\circ$, with 17 galaxies remaining.

The validity of this correction depends on the assumed geometry. If the system is instead a rotating oblate spheroid, the inferred inclination from the axis ratio will not accurately reflect the true viewing angle and the resulting correction may therefore bias the inferred rotation. However, because ECs trace recent star formation and are likely associated with the cold gas component, the assumption of a thin disk-like geometry is more physically motivated than it would be in general.

\begin{figure}
	\includegraphics[width=\columnwidth]{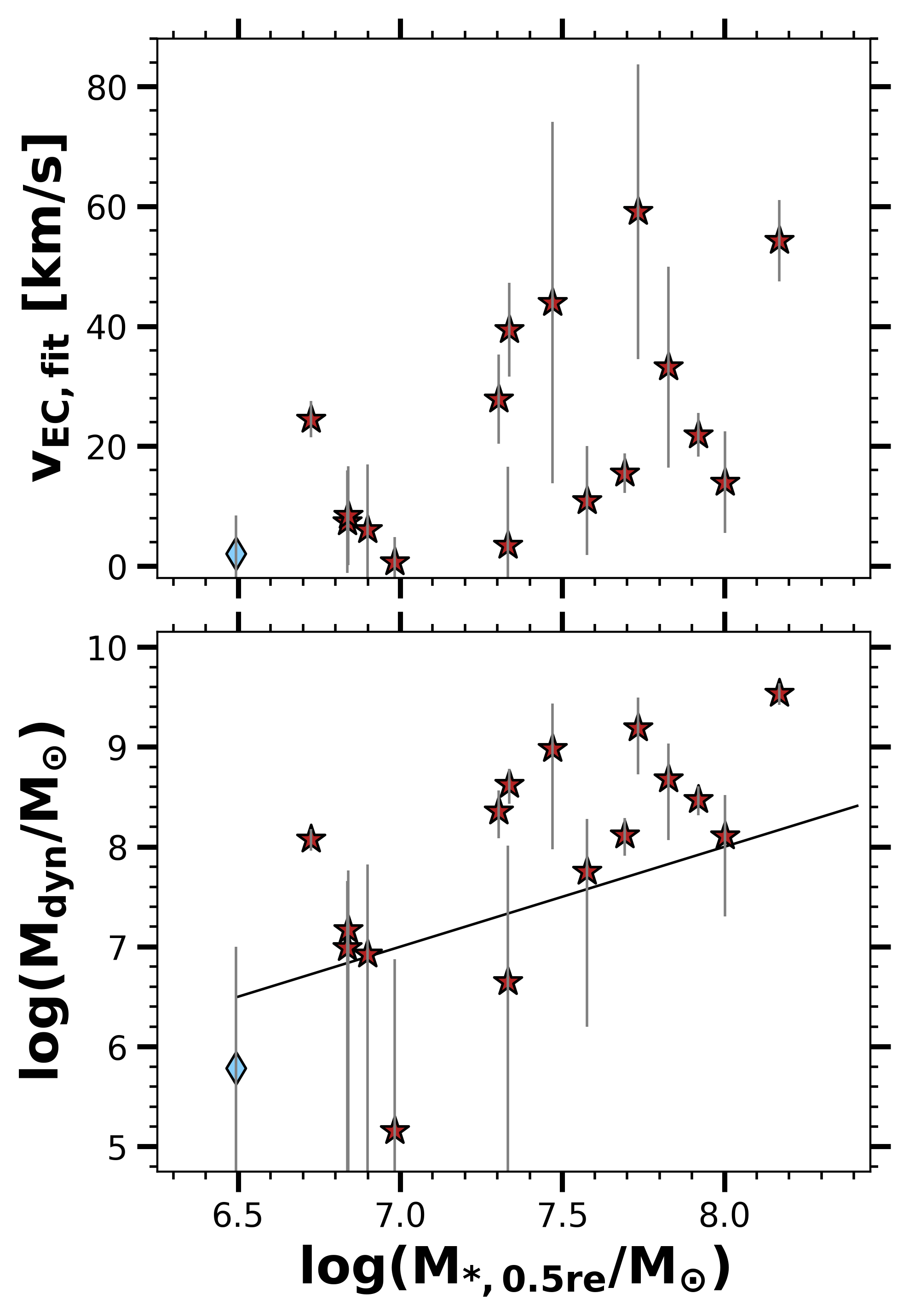}
    \caption{EC-based rotation velocities and enclosed dynamical masses as a function of color-based stellar mass \citep{roediger} estimates within 0.5 \re. In both panels, stars and diamonds represent UDGs and non-UDG dwarfs, respectively. 
    \textit{Top}: Inclination-corrected rotation velocity at 0.5 \re, obtained by fitting cluster velocities measured from a linear fit to the EC velocities along the photometric major axis. We use only ECs with velocity uncertainties smaller than 10 \kms and require at least three such ECs per galaxy. Inclinations are estimated from the photometric axis ratio of the host galaxies. 
    \textit{Bottom}: Dynamical mass enclosed within 0.5 \re, calculated from the rotation velocities. The solid black line presents the one-to-one relation. 
    }
    \label{fig:ECrot2}
\end{figure}

Figure~\ref{fig:ECrot2} presents the EC-based rotation measurements and the corresponding dynamical mass as a function of the stellar mass enclosed within 0.5\re. We estimate the enclosed stellar mass in two steps. First, we derive the total stellar mass by combining the galaxy luminosity and color with the color-dependent stellar mass-to-light ratio ($M_*$/$L_*$) prescription of \cite{roediger}. The median $g$-band stellar mass-to-light ratio is $M_*/L_* = 0.8 \pm 0.2$, where the uncertainties indicate 16th--84th percentile range.
We then scale this mass by the fraction of stellar light within 0.5\re, as determined from the S\'ersic profile.
We re-evaluate the applicability of the color-derived $M_*$/$L_*$s to galaxies in \S\ref{sec:ML_stellar}. 

We estimate the dynamical mass enclosed within 0.5\re\ as

\begin{linenomath}
    \begin{equation}
        M_{\rm dyn}(<0.5r_{\rm e}) = v_{\rm rot}^2 (0.5r_{\rm e})/(AG) \, , 
        \label{eq:dyn_mass}
    \end{equation}
\end{linenomath}
where $v_{\rm rot}$ is the rotation velocity at $0.5r_e$, $G$ is the gravitational constant, and $A$ parameterizes the correction for deviations from a spherical mass distribution. For eight nearby disk galaxies, \cite{feng} reports a mean value of $A = 1.66$, although these systems are baryon-dominated within $r_{\rm e}$.
Given the uncertain mass distribution of our galaxies, we adopt $A=1$ for simplicity. However, the true value of $A$ may be greater than unity, in which case we overestimate the dynamical masses by that factor.
In the top panel of Figure~\ref{fig:ECrot2}, we present the rotation velocity at 0.5\re, and in the bottom panel, we present the dynamical mass within 0.5\re\ derived from that velocity. 

The resulting dynamical masses generally exceed the stellar mass within the same radius. Restricting the calculation to galaxies that satisfy our UDG size criterion, we find a mean linear-scale mass ratio of $M_{\rm dyn}$/$M_*$$=9.91\pm2.84$, where the uncertainty is the standard error of the mean. This offset is consistent with the expectation that the diffuse, low-mass galaxies are dark matter dominated throughout \citep[e.g.,][]{simon2019,vdk19,forbes+21,zaritsky_2022,forbes24}.

\begin{figure}
	\includegraphics[width=\columnwidth]{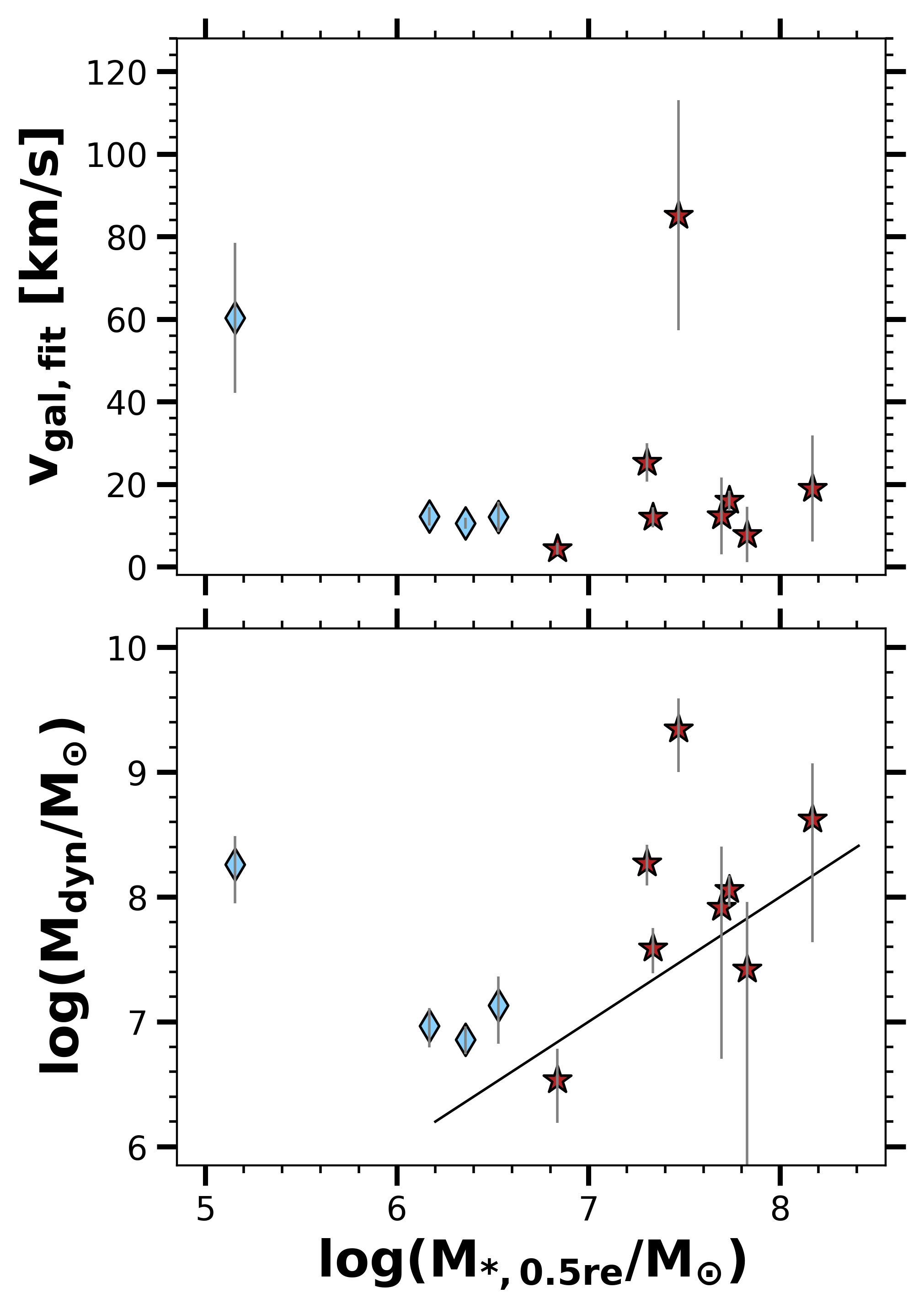}
    \caption{Fitted rotation velocity and dynamical mass obtained from the diffuse stellar light as a function of color-based stellar mass \citep{roediger} estimates within 0.5 \re, in a similar format to Figure~\ref{fig:ECrot2}. 
    \textit{Top}: Inclination-corrected rotation velocity at 0.5 \re, measured from absorption-line velocities of the diffuse stellar component. We measure the stellar velocity gradients along the photometric major axis using the procedure described in \S\ref{sec:rot_gal}. We show only the 13 galaxies for which the fitted rotation velocity exceeds its uncertainty.
    \textit{Bottom}: Dynamical mass enclosed within 0.5 \re, calculated from the fitted stellar rotation velocities.
    }
    \label{fig:Galrot}
\end{figure}

\subsubsection{Rotation derived from spatially resolved stellar kinematics}
\label{sec:rot_gal}

To characterize the large-scale kinematic structure of the underlying stellar population, we measure spatially resolved line-of-sight velocities from the stellar absorption features in diffuse galaxy light. 
We define the fitting region as the largest galaxy-shaped elliptical aperture that is fully enclosed within the field of view, adopting the axis ratio and position angle of the host galaxy. After masking all identified clusters, we use the \texttt{Powerbin} algorithm \citep{powerbin} to group the remaining spaxels into spatial bins. Because the choice of the target SNR changes the number and spatial sampling of the bins, we repeat the binning procedure for target SNR ratios from 50 to 500 in steps of 25. 

For each binning configuration, we extract binned spectra and measure the recessional velocity from absorption-line features. We retain only bins with velocity uncertainties smaller than 30 \kms. Using these bins, we perform an initial weighted least-squares fit to the velocity profile and remove outliers through 2$\sigma$ clipping. 
After removing these outliers, we retain only binning configurations that satisfy two quality criteria: more than 80\% of the pixels in the target region lie in successfully fitted bins, and at least four valid bins are used for the velocity gradient measurement. 

For each accepted configuration, we refit the cleaned velocity profile using weighted least squares. We adopt the weighted mean of the accepted slopes as the final velocity gradient. 
We reject galaxies for which the fitted rotation velocity fails to exceed the uncertainty, leaving us with 13 galaxies for which we measure the stellar rotation velocity.
We then evaluate the fitted rotation velocity at 0.5\re, and correct it for inclination, again assuming a thin disk geometry. As in the EC-based analysis, we exclude nearly face-on systems, leaving 12 galaxies with inclination-corrected stellar rotation measurements.

Figure~\ref{fig:Galrot} presents the rotation velocities measured from the diffuse stellar light and corresponding dynamical mass within 0.5\re, in a similar format to Figure~\ref{fig:ECrot2}. We note that the dynamical masses shown here are based only on the fitted rotation velocities and do not include contributions from random motions or velocity dispersion. In contrast to the EC-based measurements, these velocities trace the bulk motion of the unresolved stellar body rather than localized regions of recent star formation. Most galaxies with successful measurements show modest rotation, with $v_{\rm gal,fit} \lesssim20$ \kms, although two systems exhibit substantially higher velocities. We find no clear monotonic trend between the rotation speed and enclosed stellar mass. Nevertheless, this figure suggests that ordered motion is present in at least some diffuse stellar bodies, not only in the star-forming disk.

Seven of 12 galaxies with reliable stellar kinematic rotation measurements also have EC-based rotation measurements.
Within the relatively large uncertainties, the two rotation amplitudes are broadly comparable among blue galaxies. In the redder systems, however, the diffuse stellar component generally rotates more slowly than inferred from the EC velocities. 
This difference may indicate that the ECs trace a dynamically colder or more recently formed component that is not fully coupled to the older diffuse stellar population.
Four systems (SMDG0037442+241228, SMDG0102539+305357, SMDG0152163-044242, and SMDG2239446+180218) show signs of counter-rotation, and we examine these systems in more detail in \S\ref{sec:ma}.

\subsection{Dynamical Mass and Mass-to-Light Ratio Inside the Effective Radius}
\label{sec:ushape}

We compare our dynamical mass-to-light ratios with literature samples spanning a wide range in dynamical mass to place our galaxies in the broader context of galaxy dynamical properties. We measure our dynamical masses within 0.5\re, whereas the literature studies report values within \re, so we extrapolate our measurements to the same radius. Because this correction depends on the uncertain inner dark-matter profile of UDGs, we consider both an NFW \citep{NFW1996} cusp and a cored halo profile, with the latter having been suggested for some UDGs \citep[e.g.,][]{Gannon2020,forbes24}.

\begin{figure}
	\includegraphics[width=\columnwidth]{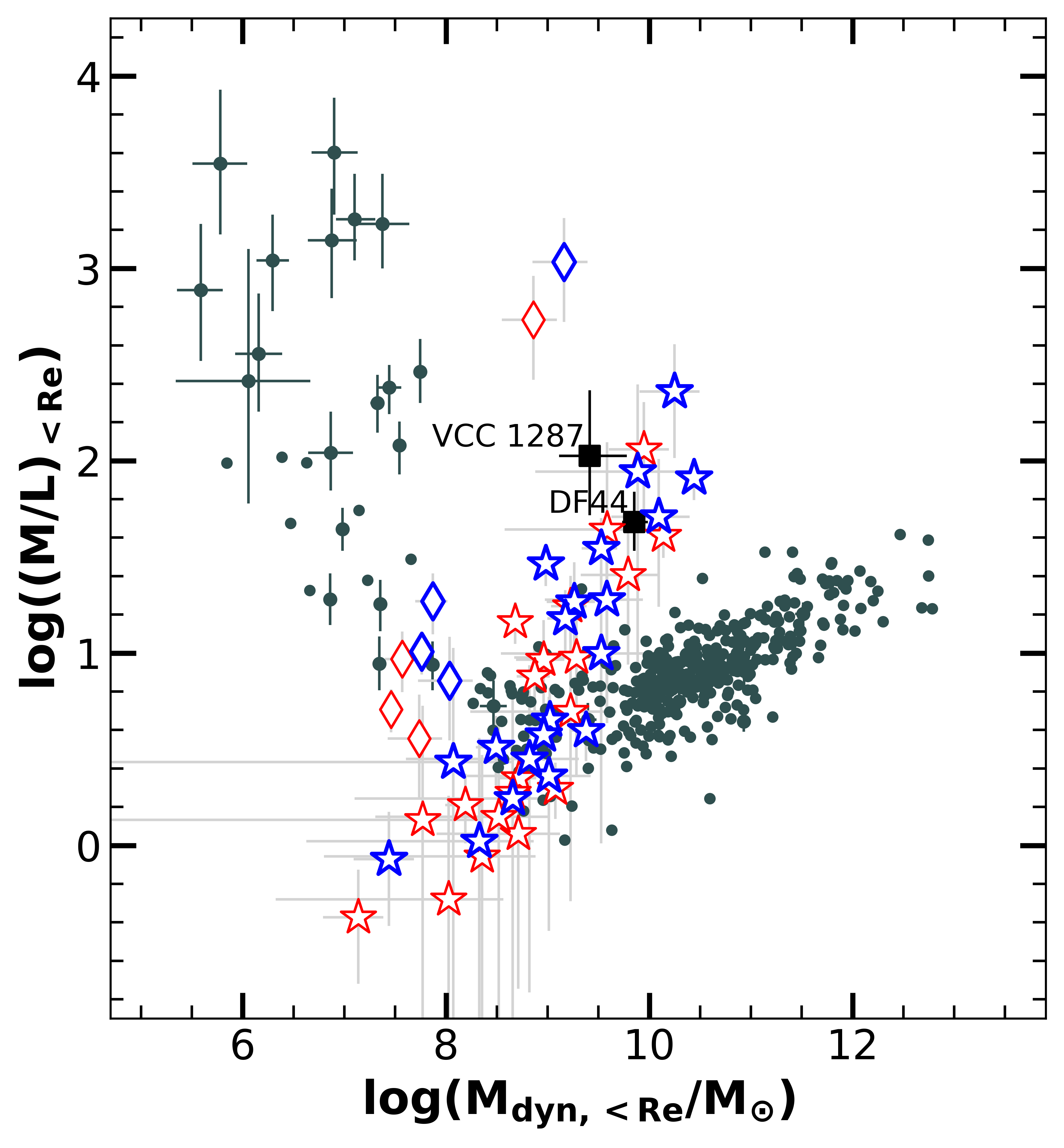}
    \caption{Dynamical M/$L_r$ ratio as a function of dynamical mass within \re. Gray symbols show literature galaxy samples spanning a broad range of galaxy mass, from dwarfs to brightest cluster galaxies \citep{Jorgensen1996,Oegerle1991,Chilingarian2008ETG,Geha2003dE,Simon2007UFD,wolf+10}. The well-studied UDGs DF44 \citep{vdk2016_df44} and VCC 1287 \citep{beasley2016} are shown as black squares. Our sample is shown with open symbols, with stars representing UDGs and diamonds representing non-UDG dwarfs. The dynamical masses of our sample were measured within 0.5\re and were therefore extrapolated to \re\ for comparison. Red and blue symbols assume an inner NFW cusp and cored halo profile, respectively. 
    }
    \label{fig:ushape}
\end{figure}

In Figure~\ref{fig:ushape}, we present that comparison, including literature samples spanning a broad range in galaxy mass, from dwarf galaxies and low-mass Local Group members to brightest cluster galaxies. 
Our sample spans a broad range in dynamical mass-to-light ratio, from baryon-dominated systems to strongly dark matter-dominated systems, similar to the known UDGs (DF44 and VCC 1287) and fills in an area of the diagram that is distinct from those covered by other galaxies. Although the inferred values depend quantitatively on the adopted halo profiles, both prescriptions preserve this broad galaxy-to-galaxy variation. The diagonal nature of the distribution of our systems in this parameter space, with the gap between our systems and the low-mass Local Group members, may reflect a selection bias because lower-mass, dark-matter-dominated analogs of our galaxies may be of even lower surface brightness and therefore escaped detection in SMUDGes. Such systems are being found \citep[e.g.,][]{Montes2024_nube,Li2025_darkgal}, but dynamical measurements for those do not yet exist. Note that an absence of dark matter within $r_e$ does not necessarily imply a global absence of dark matter.

\section{Discussion}
\label{sec:discussion}

\subsection{GC Luminosity Function}
\label{sec:GCLF}

\begin{figure}
	\includegraphics[width=\columnwidth]{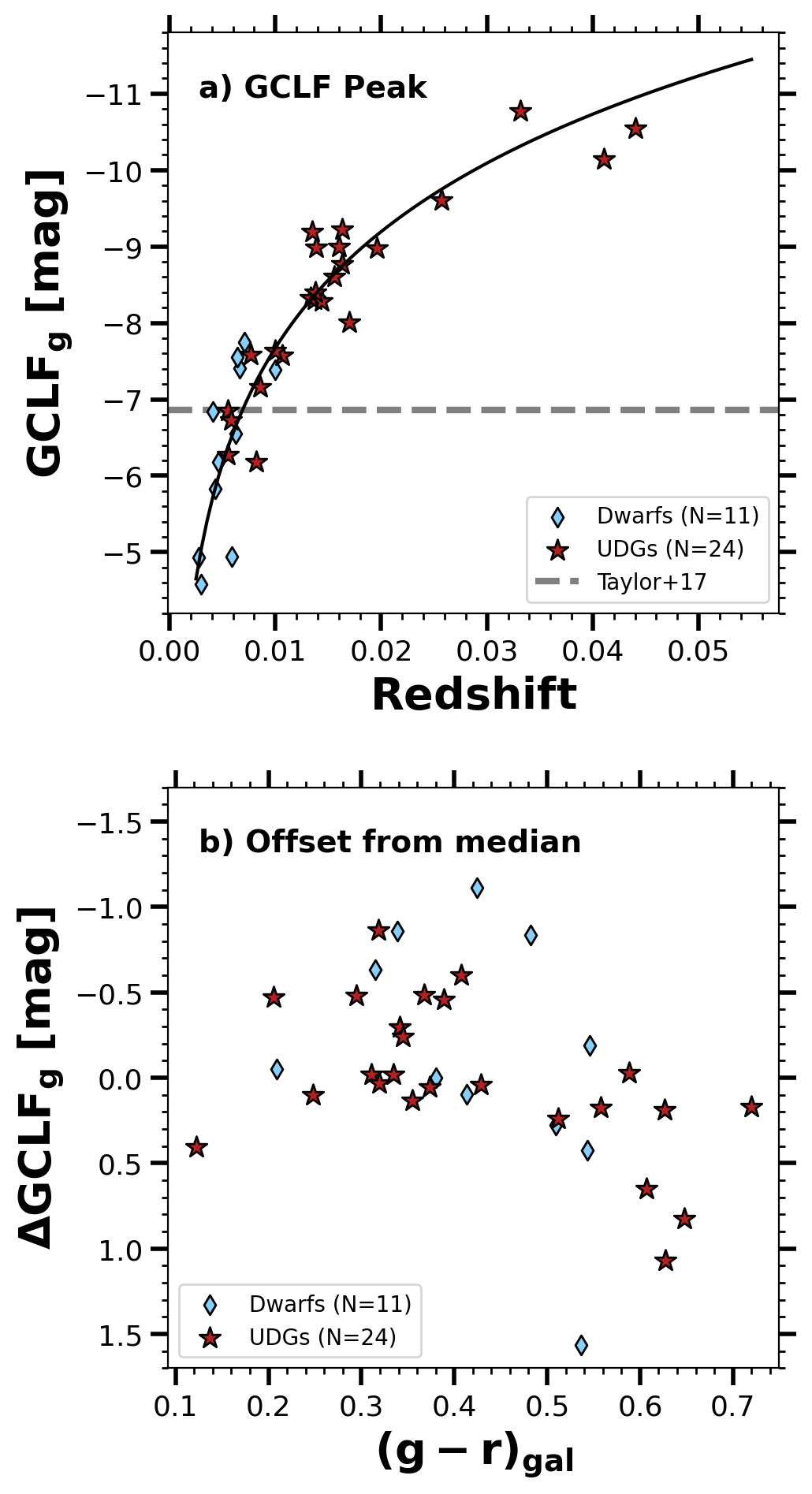}
    \caption{Comparison of the measured GCLF peak magnitudes. Panel (a): The absolute g-band GCLF peak magnitude as a function of redshift, with the symbol color indicating the integrated host galaxy (g-r) color. We include only galaxies with at least five GCs to ensure a meaningful GCLF peak estimate. The black curve shows the redshift dependence of the median apparent GCLF peak, converted to absolute magnitude, confirming that the bulk trend is due to selection effects. The gray dashed lines mark the canonical absolute g-band GCLF peak \citep{Taylor+2017}. Panel (b): Offset from the black curve as a function of the host galaxy color. Negative values indicate GCLF peaks brighter than the median value of the GCLF peaks. 
    }
    \label{fig:GCLF}
\end{figure}

The globular cluster luminosity function (GCLF) has a nearly universal turnover magnitude \citep{Jacoby+1992, McLaughlin+2005, Brodie+2006}, and similar turnover behavior has also been reported for GC populations in UDGs \citep{Saifollahi+2022,marleau2024}. In contrast, the Disco Ball shows a brighter GCLF peak, suggesting that the GCs in the Disco Ball may be younger or more luminous than ordinary old GCs drawn from the canonical GCLF. Unusually luminous GC populations have also been reported in several dark-matter-free UDGs \citep{vdk18_Gc_in_DM_free, vdk19_dmfree}, although these clusters may have formed through a pathway distinct from that of typical old GCs \citep{Shen_2021}.
Motivated by these results, we apply the same GCLF analysis used for the Disco Ball to the full KCWI sample of 43 galaxies with reliable redshifts to examine whether similarly bright GCLF peaks are present in other SMUDGes galaxies.

We use the GC candidates selected in \S\ref{sec:GC} to construct the uncorrected GCLF for each galaxy. 
Here, we reject NSC candidates because they are not part of the ordinary GC population. 
We estimate the peak (or turnover) absolute magnitude of each of these GCLFs by fitting a Gaussian profile. Because the GCLF peak magnitude is poorly constrained when only a few clusters are available, we limit this analysis to galaxies with at least five GCs. 

We compare the measured GCLF peak absolute magnitudes (GCLF$_g$) in Figure~\ref{fig:GCLF}. In panel (a), we show the absolute $g$-band GCLF peak magnitude as a function of redshift. The measured GCLF peaks show a strong redshift dependence, likely reflecting the increasing difficulty of detecting faint clusters at further distances. This explanation is confirmed by calculating the expected behavior using the median apparent magnitude of the uncorrected GCLFs and simply calculating the corresponding absolute magnitude as a function of distance (the black curve in the Figure). 

In Figure~\ref{fig:GCLF}-(b) we show the offset from this redshift-dependent median curve ($\Delta {\rm GCLF}_g$) as a function of galaxy color. 
We find that $\Delta {\rm GCLF}_g$ depends on host-galaxy color. Blue galaxies with $(g-r) < 0.5$ tend to have negative $\Delta {\rm GCLF}_g$, indicating GCLF peak magnitudes that are brighter than the median value, while red galaxies with $(g-r) > 0.5$ more often show positive $\Delta {\rm GCLF}_g$.
We find no clear distinction in this behavior between UDGs and non-UDG dwarf galaxies.

This color dependence is suggestive in light of our previous study of the Disco Ball. We found that the Disco Ball has a GCLF peak magnitude that is about 0.8 mag brighter than that of the canonical GCLF \citep{Khim+25}. We considered several possible explanations for this offset, including an error in the redshift-based distance, photometric incompleteness, a non-standard GCLF, fading of young or intermediate-age clusters, and subsequent cluster disruption. Photometric incompleteness can partly explain the offset, but attributing the full discrepancy to incompleteness alone would require a total population of about 60 GCs, roughly twice the observed number. Instead, we favored a scenario in which some of the clusters are younger and currently brighter than the ancient GCs and will fade with time. 

This fading scenario provides a natural explanation for the trend in Figure~\ref{fig:GCLF}-(b). Blue galaxies are more likely to have experienced recent or extended star formation, and their cluster populations may therefore include young or intermediate-age clusters. Our previous stellar population toy model, based on \texttt{Prospector} \citep{prospector} with the \texttt{Flexible Stellar Population Synthesis} (FSPS) stellar population models \citep{fsps1, fsps2}, showed that even a 100 Myr-old, metal-poor ([Fe/H]$=-1$) cluster, which would already lack emission lines, fades by $\sim 3.5$ mag over 10 Gyr. 
Therefore, even after we remove clusters with emission lines, the remaining cluster population in blue galaxies may still include young or intermediate-age clusters that will fade further as they age, shifting the current GCLF peak magnitudes brighter relative to that of a purely old GC population. Additionally, this result cautions against treating all photometrically selected ``GCs'' as a uniform population of old GCs, particularly in blue galaxies, because younger clusters could bias both the inferred GC counts and halo masses based on them.

\subsection{GC Counts Corrections and Inferred Halo Mass}
\label{sec:detec_limit}

We estimate total GC populations to compare GC-system richness and infer host halo masses. Because the observed counts are limited by the field of view and photometric depth, we apply radial and photometric corrections. The resulting counts are listed in Table~\ref{table:cluster_numbers}.

\subsubsection{Radial correction}

The limited field of view of KCWI, which spans $0.29$--$2.02$ \re, prevents us from detecting the full GC population. We therefore estimate a radially corrected GC population, $N_{\rm GC,rad}$, by assuming that the cluster spatial distribution follows the underlying stellar component, consistent with some previous studies of UDGs \citep{Saifollahi+2022,marleau2024}. We model the stellar light with a S\'ersic profile, adopting structural parameters from \cite{smudges2,smudges5}. Using the largest isophote that can be fully enclosed within the field of view, we estimate the fraction of the total light within that aperture, and apply the inverse of this fraction to correct the cluster counts.

Some studies instead assume that GC systems are more spatially extended than stellar light and therefore apply larger radial corrections \citep[see][for discussion]{forbes24}. Our approach yields a smaller and more conservative estimate of the GC count than those studies. 

\subsubsection{Photometric correction}

Although our previous analysis suggested that some  ``GCs'' may not be canonical old GCs, this concern should be less relevant for clusters in red-sequence galaxies, where recent cluster formation is less likely (see \S\ref{sec:GCLF}). Therefore, we restrict ourselves here to galaxies lying within $2\sigma$ of the red sequence in the color-magnitude diagram. We begin with the radially corrected GC population ($N_{\rm GC,rad}$), and then apply the correction for photometric incompleteness using the measured GCLF turnover magnitude. 

Following the approach described by \cite{Khim+25}, we treat our measured GCLF turnover as the faintest magnitude to which our observed cluster sample is complete.
We then calculate $f_{\mathrm{GCLF}}$, the fraction of the reference GCLF from \citep{Taylor+2017} that lies brighter than this limiting magnitude, and estimate the fully corrected population as
\begin{linenomath}
    \begin{equation}
        N_{\mathrm{GC,Tot}} = N_{\mathrm{GC,Rad}}/f_{\mathrm{GCLF}} \, . 
        \label{eq:GC_tot}
    \end{equation}
\end{linenomath}
\noindent
To limit the analysis to galaxies with reasonably constrained photometric corrections, we exclude systems whose measured turnover is more than 1.5 mag brighter than the \cite{Taylor+2017} peak. Such galaxies sample only the bright tail of the GCLF, making the inferred corrections highly uncertain. This cut removes galaxies at $z>0.02$ (see Figure~\ref{fig:GCLF}-(a)) and limits the photometric correction factor to $\lesssim 3.7$. 

In the top panel of Figure~\ref{fig:ngc_all}, we compare $N_{\mathrm{GC,Tot}}$ with the stellar mass of the host galaxy. The corrected counts broadly increase with stellar mass, although the scatter is large. Three of the most massive galaxies host relatively rich GC populations, while most lower-mass galaxies have substantially smaller corrected counts. UDGs, which are more massive, tend to have larger GC populations than non-UDG dwarfs.

\subsubsection{Halo Mass Inferred from GC Counts}

We infer halo masses from the $N_{\mathrm{GC,Tot}}$ using the empirical scaling relation between \ngc and halo mass from \cite{Harris2017}, as implemented in \cite{Saifollahi+2022}, while acknowledging the substantial uncertainties involved in this approach. Assuming an average GC mass of 2 $\times$ 10$^{5}$ \Msol, this relation can be expressed as
\begin{linenomath}
    \begin{equation}
        M_{\mathrm{h}} = 5.12 \times 10^{9} \times N_{\mathrm{GC}} \, \mathrm{M_\odot}. 
        \label{eq:scaling_relation}
    \end{equation}
\end{linenomath}
\noindent

\begin{figure}
	\includegraphics[width=\columnwidth]{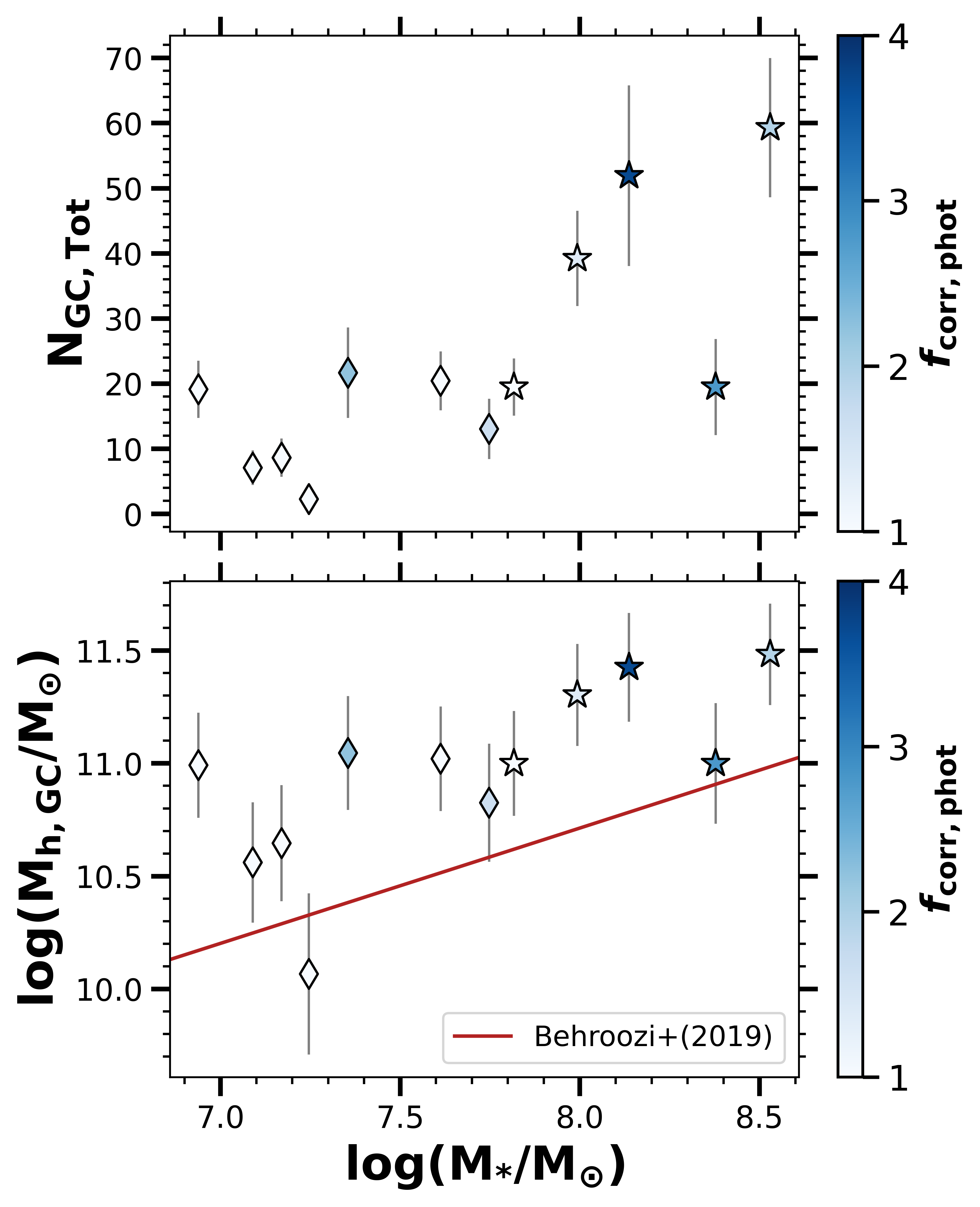}
    \caption{\textit{Top}: Radially and photometrically corrected \ngc ($N_{\rm GC,Tot}$) as a function of the stellar mass of the galaxies lying within $2\sigma$ of the red sequence. The color shows the photometric number correction factor ($=$ boost factor). Stars and diamonds represent UDGs and non-UDG dwarfs, respectively. The corrected counts generally increase with stellar mass.
    UDGs seem to host larger GC populations, but this trend may primarily reflect their higher stellar mass. \textit{Bottom}: Halo mass inferred from $N_{\rm GC,Tot}$. The red curve represents the stellar mass--halo mass relation of \cite{Behroozi2019}. The GC-based halo masses lie systematically above this relation.
    }
    \label{fig:ngc_all}
\end{figure}

The bottom panel of Figure~\ref{fig:ngc_all} shows the resulting GC-based halo masses as a function of stellar mass. At fixed stellar mass, the inferred halo masses generally lie above the stellar mass--halo mass relation of \cite{Behroozi2019}. This offset implies that the GC scaling relation assigns comparatively massive dark matter halos to many galaxies in the sample. This result is consistent with previous findings that UDGs tend to lie below the canonical stellar mass--halo mass relation \citep{smudges5}, suggesting that they form stars inefficiently relative to their halo masses. However, the possible presence of intermediate-age clusters, together with uncertainties in the radial and photometric corrections, limits the precision of the halo mass estimates for individual galaxies.

\subsection{Stellar Mass to Light Ratio}
\label{sec:ML_stellar}

In \S\ref{sec:rot_ec}, we estimate the enclosed stellar mass using a color-dependent stellar $M$/$L$ ratio \citep{roediger}. This approach provides a homogeneous estimate for the full sample and is commonly used in the literature \citep[e.g.,][]{smudges5,roman17b,Chamba2019}, but it may hide subtle differences related to the details of the star formation histories, the stellar initial mass function, or metallicity of these galaxies. For example, in bursty dwarf galaxies, the color varies significantly with time and may be a blunt tool with which to determine the stellar mass.  

To address this concern, we compare these color-based stellar masses with independent estimates obtained from detailed spectral energy distribution (SED) fitting. \cite{SMUDGes9} estimated total stellar masses for SMUDGes galaxies by reconstructing the observed SEDs with \texttt{PROSPECTOR}. In brief, they applied SED fitting to galaxies with acceptable photometry in all three $g$, $r$, and $z$ bands, at least one usable GALEX UV and one WISE IR measurement, and an available redshift estimate. After correcting the multi-band photometry for Galactic extinction, they modeled the SEDs with FSPS stellar population models and a nonparametric star formation history. The total stellar mass was treated as one of the free parameters in the fit.

\begin{figure}
	\includegraphics[width=\columnwidth]{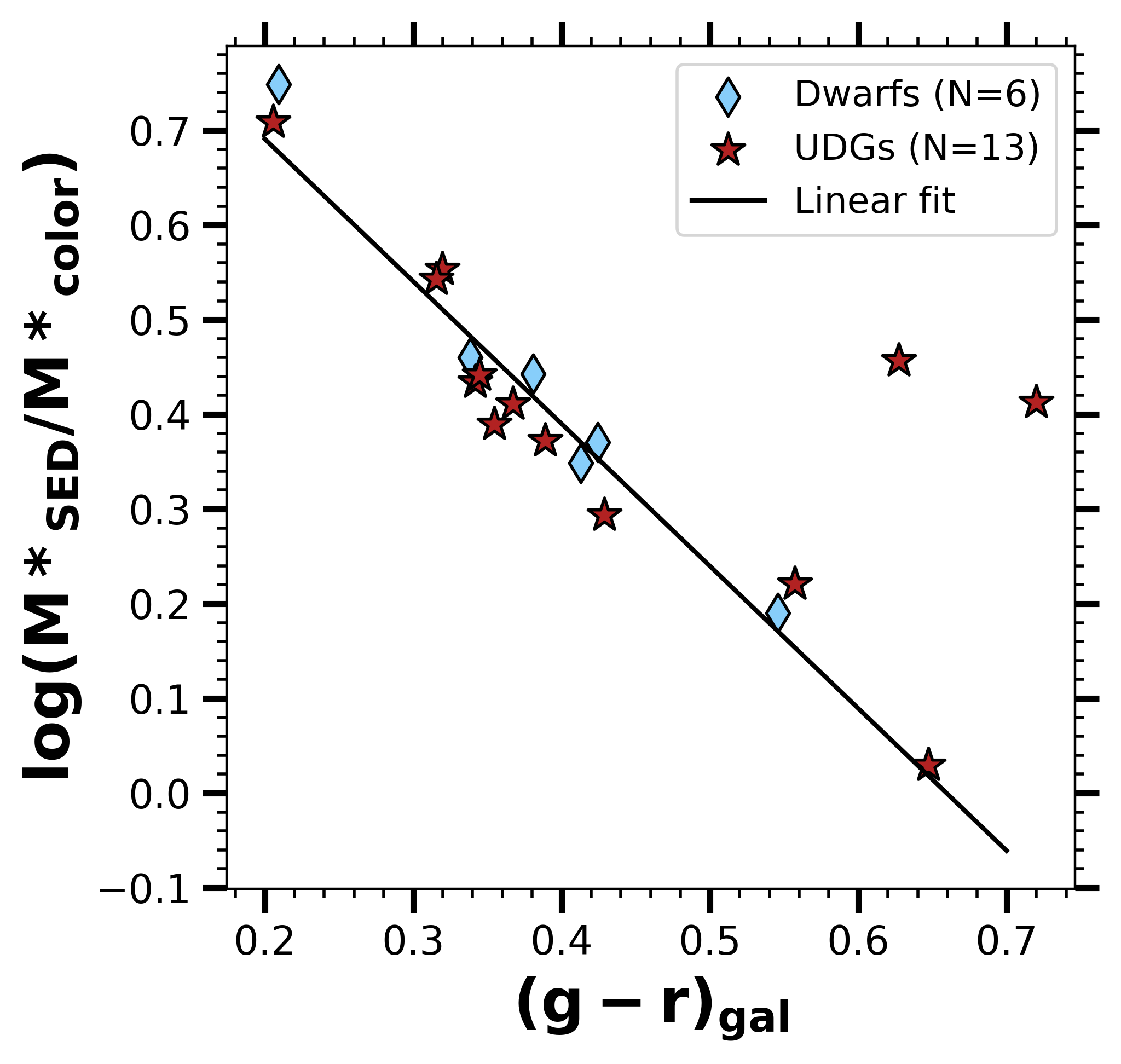}
    \caption{Comparison of SED-based and color-based stellar masses as a function of galaxy color. We show the 19 galaxies with available SED-based measurements. The offset between the two mass estimates depends strongly on galaxy color. The solid line shows a linear fit after excluding the two red outliers, both of which have high inferred dust content from the SED analysis.
    }
    \label{fig:mlratio}
\end{figure}

Among the 44 galaxies in our sample, 19 satisfy the requirements for analysis and therefore have SED-based stellar mass estimates. Figure~\ref{fig:mlratio} compares the SED-based and color-based stellar masses as a function of galaxy color. The SED-based masses are generally larger than the color-based estimates, with the offset increasing toward bluer color. We exclude two red galaxies that are clear outliers to the linear color-dependent relation; both have high inferred dust contents in the SED analysis and so are questionable. The remaining 17 galaxies follow a well-defined relation, with UDGs and non-UDG dwarfs occupying the same sequence. The systematic offset therefore appears to depend primarily on galaxy color rather than on whether a galaxy satisfies the UDG size criterion.

We use these 17 galaxies to derive an empirical color-dependent correction and apply this relation to estimate SED-calibrated stellar masses for the full sample. After this correction, the median $g$-band stellar mass-to-light ratio is $M_*/L_* = 2.3^{+0.4}_{-0.2}$, where the uncertainties indicate 16th--84th percentile range. The larger stellar masses reduce the median dynamical-to-stellar mass ratio to $M_{\rm dyn}$/$M_*$$=4.02\pm1.02$. 
Importantly, the revised stellar masses do not remove the large galaxy-to-galaxy variation in $M_{\rm dyn}$/$M_*$. The corrected ratios still span a broad range, including systems that appear baryon-dominated within 0.5\re\ and others that retain a large dark matter fraction over the same radial range.

\subsection{Kinematic Misalignment between ECs and Diffuse Stellar Light}
\label{sec:ma}

We previously noted that the ECs and diffuse stellar light do not always have aligned angular momenta. This difference may reflect the distinct components traced by the two measurements: the ECs are associated with recent star formation and may follow the cold gas from which they formed, whereas diffuse stellar light traces the bulk motion of the older stellar body.
Such differences are not unique to our sample. Previous studies of UDGs and dwarf galaxies have reported offsets between optical and \hi\ position angles \citep[e.g.,][]{Gault2021, Pina2022}, as well as direct kinematic misalignment between their stellar and gaseous components \citep[e.g.,][]{Graham2017,2020A&A...634A..10H}. 

If the opposing fitted gradients reflect a genuine difference in rotational direction, they may indicate that the galaxies acquired gas after the formation of the old stellar body. Studies of more massive galaxies have proposed several scenarios for star-gas misalignment and counter-rotation. Gas-rich mergers \citep[e.g.,][]{Balcells1990,Barnes2002}, interactions with nearby galaxies \citep[e.g.,][]{Bournaud2003,DeRijcke2004}, and continuous accretion through cosmic filaments \citep[e.g.,][]{Thakar1996,Brook2008} can introduce gas whose angular momentum differs from that of the pre-existing stellar body. Cosmological simulations also show that mergers, interactions, and externally accreted gas can produce star-gas misalignments and counter-rotating structures \citep{Starkenburg2019,Duckworth+20,Khim_MA,Khim_MA2}. Mergers may be especially relevant in the context of UDGs, as they have also been proposed as a formation channel for at least some UDGs \citep[e.g.,][]{Fielder2024_UDGmerger}.

Among the seven galaxies with reliable measurements of both components, four show EC velocity gradients with the opposite sign from those measured from the diffuse stellar light. We do not classify these systems as definitive counter-rotators, because both measurements (EC and stellar) rely on sparse spatial sampling and the results are sensitive to the locations of individual ECs or stellar-light bins. This limitation is clear for SMDG0037442+241228, which has a strongly negative EC rotation velocity gradient ($59\pm24$ \kms), but has ECs detected only on one side of the galaxy.

\begin{figure}
	\includegraphics[width=\columnwidth]{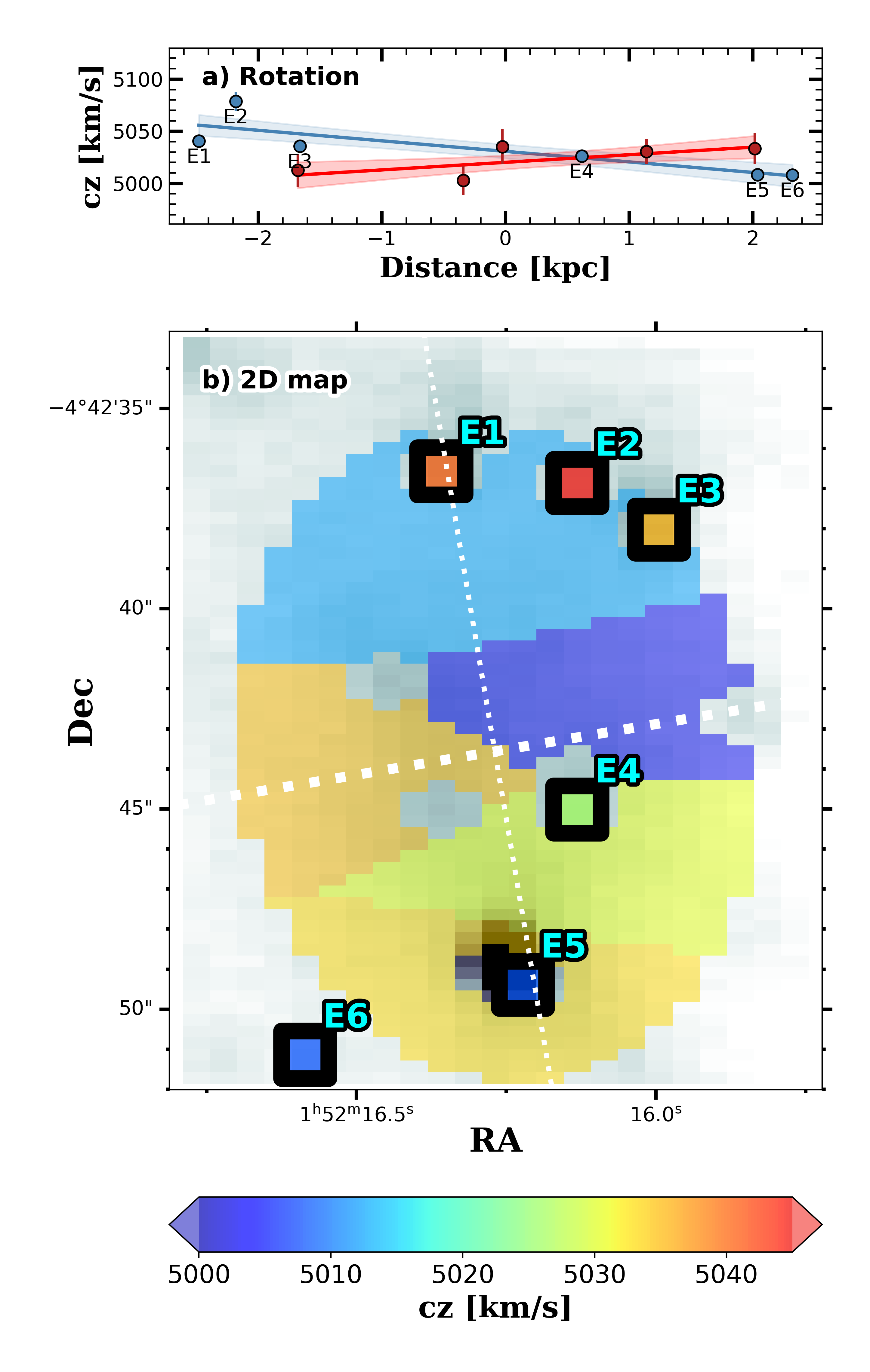}
    \caption{Example of a galaxy in which the ECs and diffuse stellar light show opposite fitted velocity gradients.
    \textit{Panel (a)}: Recessional velocities of the diffuse stellar light bins (red) and ECs (blue) as a function of projected distance along the photometric major axis. Solid lines and shaded areas show the best-fit linear gradient and their 1$\sigma$ uncertainties.
    \textit{Panel (b)}: Spatial distribution of the measured velocities. The thin and thick white dotted lines indicate the photometric major and minor axes, respectively.
    }
    \label{fig:MA}
\end{figure}

In Figure~\ref{fig:MA}, we show our most convincing case for counter-rotation. The diffuse stellar-light bins show a mild positive velocity gradient along the photometric major axis, whereas the ECs show a negative gradient over a similar projected baseline. The two-dimensional velocity map illustrates the same behavior. Nevertheless, the small number of ECs and reliable stellar-light bins limits the robustness of the fitted gradients. 

The possible misalignment is relevant to our geometric correction. We estimate a single inclination angle from the photometric axis ratio of each host galaxy and apply the same thin-disk correction to both the EC-based and diffuse-light-based rotation measurements. If ECs and the diffuse stellar body trace different angular momentum components, this common inclination correction might not be appropriate for both. We therefore interpret the inclination-corrected velocities in these systems with some caution.

\subsection{Searching for Disco Ball Analogs in the Sample}
\label{sec:analogs}

Motivated by the results and discussions above, we search for Disco Ball analogs within the KCWI sample. We characterize the Disco Ball by three key properties: (1) its green-valley location in the color-magnitude diagram, (2) its rich and diverse high surface brightness stellar components, including GCs, ECs, and an NSC, and (3) its measurable and mutually consistent rotation in both ECs and diffuse stellar light \citep{Khim+25}.

We find no exact counterpart that satisfies all three criteria. Seven galaxies have measurable rotation from both ECs and diffuse stellar light, but they are substantially bluer ($g-r<0.43$) than the Disco Ball ($g-r=0.51$). In addition, none of these kinematic analogs hosts an NSC candidate within 0.1\re. 

Nevertheless, we identify SMDG0152163-044242 as the closest analog to the Disco Ball in our sample. This galaxy shows measurable rotation from both ECs and diffuse stellar light and hosts an NSC candidate within 0.2\re. As discussed above, the 0.1\re\ cut is intentionally conservative and may miss plausible NSC candidates in our sample due to uncertainties in the photometric center, the clumpy central structure and the limited spatial resolution of the ground-based data (see \S\ref{sec:NSC}). Under this definition, SMDG0152163-044242 overlaps with the Disco Ball in both kinematic behavior and possible NSC presence, although it is substantially bluer ($g-r=0.40$) than the Disco Ball. 

Other galaxies provide more limited comparisons. In terms of compact-source content, SMDG0015089-031837 is also notable because it hosts an NSC candidate as well as multiple ECs and GCs. This galaxy shows measurable EC rotation, but its greater distance ($z\sim0.02$) and low S/N prevent us from testing whether the diffuse stellar component shares the same rotational signature. Finally, SMDG2337078+001240 lies close to the red sequence and hosts one EC, but lacks both the richer cluster population and the measurable rotation seen in the Disco Ball.

These comparisons show that none of the individual properties of the Disco Ball is unique within the KCWI sample. Rotation is not unique among galaxies with recent star formation, some systems host NSC candidates, and others contain a diverse population of clusters. However, the Disco Ball remains unusual because it combines these properties in a relatively red transitional system. In this sense, the Disco Ball does not appear to represent an entirely distinct class of galaxy, but rather an uncommon conjunction of features that usually appear separately in our sample.

\subsection{Diversity of Dark Matter Content in UDGs}
\label{sec:diversity_dm}

Early dynamical studies of UDGs identified systems such as VCC 1287 and DF44 with unusually large dynamical mass-to-light ratios, motivating the interpretation that at least some UDGs host disproportionately massive dark-matter halos. Our larger field sample presents a more heterogeneous picture. While several galaxies occupy the high mass-to-light ratio regime represented by these well-studied UDGs, others have dynamical masses within \re\ that are only modestly larger than their enclosed stellar masses (Figures~\ref{fig:ECrot2}--\ref{fig:ushape}). High inner dark-matter fractions therefore do not appear to be a universal property of UDGs.

This broad distribution further supports the view that galaxies with similarly diffuse stellar structure can have substantially different halo mass profiles. In particular, the UDGs in our sample do not appear to form a single dynamically homogeneous population. Instead, the distribution may hint at populations with relatively low and high inner dark-matter fractions. 

Previous work has shown that GC-rich UDGs can also have large dynamical masses, and GC abundance has been used as an empirical tracer of total halo mass. A correspondence between GC richness and inner dynamical mass fraction would therefore provide a possible connection between cluster population and dark-matter content \citep[e.g.,][]{Gannon2022,forbes24}. 
However, we cannot directly test this connection with the present data. Most galaxies for which we obtain dynamical measurements are blue in color, whereas we restrict our total GC population estimates to galaxies near the red sequence to minimize contamination from young or intermediate-age clusters. 

We also caution that our dynamical measurements constrain the mass only within the central regions of the galaxies. The conversion from our measurement at 0.5\re\ to masses within \re\  depends on the assumed inner halo profile, and extrapolation to the total halo mass would introduce substantially greater model dependence. We therefore interpret the present results primarily as diversity in the enclosed mass budget rather than as precise measurements of total halo masses of individual galaxies.

\subsection{Diversity of UDGs and Formation Scenarios}
\label{sec:formation}

The broad range of properties observed in our field UDG sample suggests that these galaxies may appear structurally similar while differing in their recent star formation activities, inner density profiles, and total halo masses. This diversity motivates a comparison with commonly proposed UDG formation scenarios. 

Because we selected our sample outside of cluster environments, mechanisms associated with strong cluster processing, such as tidal heating, tidal stripping, and ram-pressure stripping, are unlikely to provide a dominant formation pathway for the sample as a whole. Such processes may nevertheless have affected individual galaxies through past interactions or pre-processing in group environments.

Some of our UDGs may be consistent with the feedback-driven expansion, or ``puffed-up dwarf" scenario. For example, \cite{DiCintio2017} identified UDG analogs in the NIHAO simulations with halo mass of approximately $10^{10}$--$10^{11}$\Msol, a broad range of colors, substantial \hi~ reservoirs, and dark matter density profiles shallower than the standard NFW profile. These properties are broadly compatible with a subset of galaxies in our sample. However, our observations do not directly constrain the bursty star formation histories or strong gas outflows that drive expansion in these models. 

The ``failed galaxy'' interpretation may also apply to a subset of our sample, particularly GC-rich, red UDGs with inferred halo mass larger than expected for their stellar masses. In this scenario, star formation proceeded inefficiently despite the galaxy residing in relatively massive halos. However, we currently lack systems for which both reliable GC-based total halo masses and rotation-based inner dynamical masses can be measured simultaneously. We therefore cannot directly test whether galaxies with relatively massive halos also have correspondingly large central dark matter contributions. At the other extreme, several galaxies have dynamical masses within 0.5\re\  that are comparable to their enclosed stellar masses. These systems may be related to proposed dark-matter-deficient formation channels \citep[e.g.,][]{Ogiya2022,vdk2022Nature}, although the present measurements constrain only the inner mass distribution and do not establish that the galaxies possess unusually low total halo masses.

Overall, our observed diversity in our sample favors a picture in which multiple formation and evolutionary pathways contribute to the field UDG population. Similar diffuse structures may therefore arise from different combinations of baryonic evolution and dark matter halo properties.

\section{Summary}
\label{sec:summary}

We use spatially resolved KCWI spectroscopy of 44 LSB galaxies selected from the SMUDGes survey to investigate the variation among populations of compact stellar sources and the enclosed dark matter distributions. The sample includes 30 galaxies that satisfy the UDG size criterion and span the red sequence, green valley, and blue cloud regions of the galaxy color-magnitude diagram. Although these galaxies appear largely smooth in wide-field survey imaging, we detect at least one compact high-surface-brightness feature in every galaxy.
The detected populations vary substantially, from galaxies hosting only clusters similar to globular clusters (GCs) to galaxies containing multiple emission-line clusters (ECs) and/or nuclear star clusters (NSCs). 
We examine how these cluster populations vary with host-galaxy properties and what the EC kinematics say about the variation in dark matter content. 
Our principal results are as follows.

\medskip
\noindent
$\bullet$ \textbf{EC populations are ubiquitous in field LSB galaxies and reflect significant variations in recent star formation activity across host-galaxy color.} We identify at least one EC in 24 of the 43 galaxies with reliable redshifts. Galaxies near the red sequence generally host no ECs, whereas nearly all blue cloud galaxies host multiple ECs. 
A smooth LSB appearance does not imply an internally uniform or quiescent system. The incidence and abundance of ECs, together with the diffuse nebular emission, reveal substantial variation in both the level and spatial organization of recent star formation, confirming that galaxies selected by similar size and surface brightness can occupy different evolutionary states \citep{loraine,SMUDGes9}.

\medskip
\noindent
$\bullet$ \textbf{Clusters without emission lines (continuum clusters or CCs) are not exclusively old GCs.} We identify at least one CC candidate in all but one galaxy across the full range of host-galaxy colors. We interpret our finding that blue galaxies have a brighter apparent CC luminosity function peak than red galaxies to mean that at least some CCs in blue galaxies are relatively younger and more luminous than typical old GCs and will fade as they age. Galaxies with similar diffuse morphologies can therefore host cluster systems formed at different epochs, complicating the interpretation of photometrically selected GC populations as uniform tracers of halo mass.

\medskip
\noindent
$\bullet$ \textbf{NSC candidates occur in galaxies with diverse global properties.} Three galaxies host NSC candidates within 0.1\re\ of their photometric centers. These candidates are at least 0.5 mag brighter than the other clusters within \re\ and have velocities consistent with those of their host galaxies. The brightest candidate shows narrow [O{\sc II}], H$\beta$, and [O{\sc III}] emission, raising the possibility of in-situ star formation or weak nuclear activity. Expanding the search radius to 0.2\re\ identifies six additional NSC candidates, although three of them have spectra with insufficient SNR to confirm their association with the host galaxies. The candidate hosts span a range of colors and cluster populations, while their differing emission-line properties suggest that their nuclear stellar components may not share the same evolutionary state.

\medskip
\noindent
$\bullet$ \textbf{Rotation traced by ECs and diffuse stellar light varies in strength and alignment across the sample.} We measure EC-based major axis velocity gradients for 17 galaxies with at least three ECs whose velocity uncertainties are smaller than 10 \kms. Most inferred rotation velocities are below 30 \kms\ at 0.5\re. We independently measure stellar velocity gradients from the diffuse galaxy light and obtain rotation measurements for 12 galaxies, with most fitted velocities below 20 \kms\ at 0.5\re. Seven galaxies have reliable rotation measurements from both tracers, and four show EC and stellar velocity gradients with opposite signs. We have less confidence in the stellar rotation measurements, and sparse spatial sampling prevents us from concluding that these systems show definitive counter-rotation. Further investigation is warranted.

\medskip
\noindent
$\bullet$ \textbf{The inferred dark matter contribution at small radii spans a broad range.} The rotation-based dynamical masses within 0.5\re\ generally exceed the enclosed stellar masses, with mean $M_{\rm dyn}$/$M_*$ ratios of $9.91\pm2.84$ for color-based stellar masses and $4.02\pm1.02$ for the SED-based stellar mass. On average, these galaxies are dark matter dominated even within 0.5\re. However, individual ratios range from systems consistent with a lack of dark matter within 0.5\re\ to systems with fully dominant dark matter contributions.

\medskip
\noindent
$\bullet$ \textbf{GC-based halo mass estimates suggest that all lie within dominant dark matter halos}.
For galaxies near the red sequence, we estimate total halo masses from the radially and photometrically corrected GC populations. The inferred halo masses generally increase with stellar mass, but lie above published stellar mass--halo mass relations. Uncertainties in cluster ages and in the radial and photometric corrections limit the reliability of these estimates for individual galaxies. 
However, at face value these measurements either indicate that galaxies with similar diffuse morphologies can have substantially different inner mass distributions for a similar halo mass or that the GC-halo mass is strongly violated in some galaxies. 

\bigskip
Overall, our results show that galaxies selected by similar low-surface-brightness and size criteria span a broad range of cluster populations, recent star formation, internal kinematics, and inferred dark matter content. The variation in cluster populations and luminosities points to different star formation histories, while kinematics and a broad range of dark matter contributions suggest diverse assembly histories. These properties vary both individually and in combination, indicating that the UDG classification encompasses galaxies in different evolutionary states and likely formed through multiple evolutionary pathways. Similar diffuse structures can therefore arise from different combinations of baryonic evolution and dark matter halo properties.

\begin{acknowledgments}
The authors acknowledge financial support from NSF AST-1713841, AST-2006785, and AST-2510821 as well as from NASA grant 22-ADAP22-0011. An allocation of computer time from the UA Research Computing High Performance Computing (HPC) at the University of Arizona and the prompt assistance of the associated computer support group is gratefully acknowledged.
This research utilizes images from the Dark Energy Camera Legacy Survey (DECaLS; Proposal ID 2014B-0404; PIs: David Schlegel and Arjun Dey). Full acknowledgment at https://www.legacysurvey.org/acknowledgment/.
\end{acknowledgments}




%
\facilities{KCWI}

\software{
\texttt{Astropy }             \citep{astropy1, astropy2},
\texttt{dustmaps}            \citep{green},
\texttt{GALFIT }              \citep{peng},
\texttt{Matplotlib }          \citep{matplotlib},
\texttt{NumPy }               \citep{numpy},
\texttt{pandas }              \citep{pandas},
\texttt{scikit-learn }         \citep{sklearn},
\texttt{SciPy}                \citep{scipy1, scipy2},
\texttt{Source Extractor Python library}                \citep{sep},
\texttt{Source Extractor}     \citep{bertin},
\texttt{PYPHOT}             \citep{zenodopyphot}
}





\bibliography{references}{}
\bibliographystyle{aasjournalv7}



\end{document}